\documentclass[envcountsame,envcountsect,undated,nolinenumbers]{lnthi}

\usepackage{epsf}

\title{Small Uncolored and Colored Choice Dictionaries}

\author{Torben Hagerup}
\institute{Institut f\"ur Informatik, Universit\"at Augsburg, 86135
Augsburg, Germany
\email{hagerup@informatik.uni-augsburg.de}}

\setpagewiselinenumbers

\def\Tvn#1{\hbox{\textit{#1\/}}}
\def\Tfloor#1{\lfloor #1\rfloor}
\def\Tceil#1{\lceil #1\rceil}
\def\TbbbN{\mathbb{N}}
\def\wik{\widetilde{\mu}}

\begin{document}
\overfullrule=5pt
\maketitle{}

\begin{abstract}
A choice dictionary is a
data structure that can be initialized
with a parameter $n\in\TbbbN=\{1,2,\ldots\}$
and subsequently maintains an initially empty subset $S$
of $\{1,\ldots,n\}$ under insertion, deletion, membership queries
and an operation \Tvn{choice} that returns
an arbitrary element of $S$.
The choice dictionary is fundamental
in space-efficient computing and has numerous applications.
The best previous choice dictionary can be initialized
with $n$ and a second parameter $t\in\TbbbN$ in constant time
and subsequently executes all
operations in $O(t)$ time and occupies
$n+O(n({t/w})^t+\log n)$ bits on a word
RAM with a word length of $w=\Omega(\log n)$ bits.
We describe a new choice dictionary that,
following a constant-time initialization,
executes all operations in constant time
and, in addition to the space
needed to store the integer~$n$, occupies only $n+1$ bits,
which is shown to be optimal if $w=o(n)$.

A generalization of the choice dictionary
called a colored choice dictionary
is initialized with a second parameter $c\in\TbbbN$
in addition to $n$ and subsequently
maintains a semipartition $(S_0,\ldots,S_{c-1})$
of $U=\{1,\ldots,n\}$, i.e., a sequence of $c$
(possibly empty) disjoint subsets of $U$
whose union is~$U$, under the operations
$\Tvn{setcolor}(j,\ell)$ ($j\in\{0,\ldots,c-1\}$
and $\ell\in U$), which moves $\ell$ from its
current subset to $S_j$, $\Tvn{color}(\ell)$
($\ell\in U$), which returns the unique
$j\in\{0,\ldots,c-1\}$ with $\ell\in S_j$,
and $\Tvn{choice}(j)$ ($j\in\{0,\ldots,c-1\}$),
which returns an arbitrary element of~$S_j$.
We describe new colored choice dictionaries
that, if initialized with constant~$c$,
execute \Tvn{setcolor},
\Tvn{color} and \Tvn{choice} in constant time
and occupy $n\log_2\!c+1$ bits plus the
space needed to store~$n$ if $c$ is
a power of~2, and at most $n\log_2\!c+n^\epsilon$
bits in general, for arbitrary fixed $\epsilon>0$.
We also
study the possibility of iterating over the set $S$
or over $S_j$ for given $j\in\{0,\ldots,c-1\}$.
This allows us to derive new
results for space-efficient breadth-first search (BFS).
On a directed or undirected graph with $n$ vertices
and $m$ edges, we can carry out a BFS either in
$O(n\log n+m\log\log n)$ time with
$n\log_2\!3+O((\log n)^2+1)$ bits
of working memory or in
$O(n\log n+m)$ time with at most $n\log_2\!3+n^\epsilon$ bits
for arbitrary fixed $\epsilon>0$.
The best previous algorithm is faster, $O(n+m)$ time, but
needs more space, $n\log_2\!3+O({n/{(\log n)^t}})$ bits
for arbitrary fixed $t\in\TbbbN$.

\smallskip
\noindent
{\bf Keywords.} Data structures, space efficiency,
choice dictionaries, bounded universes,\break
constant-time initialization, breadth-first search (BFS).
\end{abstract}

\pagestyle{plain}
\thispagestyle{plain}

\section{Introduction}
\label{sec:intro}
Concurrently with the extreme growth in the
size of data sets,
there is a trend towards the complete
data not being stored locally on a user's computer.
The data may be provided by a remote server,
it may be in a ``cloud'', or it may even
exist only in the form of an interface that can
answer queries.
In such scenarios, and also if the ``computer''
is in fact a small mobile device, it may be
important to use only (relatively) little
local memory and small data structures.
Space efficiency may be even more generally
beneficial in view of the ubiquitous
memory hierarchies that operate according to
the tradeoff ``the bigger, the slower''.
This paper deals with one particular class of
space-efficient data structures
and their applications.

Following similar earlier
definitions~\cite{BriT93,ElmHK15} and concurrently
with that of \cite{BanCR16},
the choice-dic\-tion\-ary data type was introduced
by Hagerup and Kammer~\cite{HagK16} as a basic tool in
space-efficient computing and
is known to have
numerous applications
\cite{BanCR16,ElmHK15,HagK16,HagKL17,KamKL16}.
Its precise characterization is as follows:

\begin{definition}
\label{def:choice}
A \emph{choice dictionary} is a data type that
can be initialized with an arbitrary integer
$n\in\TbbbN=\{1,2,\ldots\}$,
subsequently maintains
an initially empty subset $S$ of $U=\{1,\ldots,n\}$
and supports the following operations,
whose preconditions are indicated in
parentheses:

\begin{tabbing}
\quad\=\hskip 2.4cm\=\hskip 2cm\=\kill
\>$\Tvn{insert}(\ell)$\>($\ell\in U$):\>
Replaces $S$ by $S\cup\{\ell\}$.\\
\>$\Tvn{delete}(\ell)$\>($\ell\in U$):\>
Replaces $S$ by $S\setminus\{\ell\}$.\\
\>$\Tvn{contains}(\ell)$\>($\ell\in U$):\>
Returns 1 if $\ell\in S$, 0 otherwise.\\
\>$\Tvn{choice}$:\>
\>Returns an (arbitrary) element of $S$
if $S\not=\emptyset$, 0 otherwise.
\end{tabbing}
\end{definition}

Informally, a choice dictionary can be understood
as a bit vector with the additional operations
``clear all'' (through renewed initialization)
and ``locate a~\texttt{1}''.
As is common and convenient, we use the term
``choice dictionary'' also to denote data structures
that implement the choice-dictionary data type.
Following the initialization of a choice dictionary
$D$ with an integer~$n$,
we call (the constant) $n$ the \emph{universe size}
of $D$ and (the variable) $S$ its
\emph{client set}.
If a choice dictionary $D$ can operate only if
given access to $n$ (stored outside of~$D$),
we say that $D$ is \emph{externally sized}.
Otherwise, for emphasis, we may call $D$
\emph{self-contained}.

Our model of computation is a word RAM~\cite{AngV79,Hag98}
with a word length of $w\in\TbbbN$ bits, where we assume that $w$ is
large enough to allow all memory words in use to be addressed.
As part of ensuring this,
in the discussion of a choice dictionary
with universe size~$n$ we always assume that $w\ge\log_2\! n$.
The word RAM has constant-time operations
for addition, subtraction and multiplication
modulo $2^w$, division with truncation
($(x,y)\mapsto\Tfloor{{x/y}}$ for $y>0$),
left shift modulo $2^w$
($(x,y)\mapsto (x\ll y)\bmod 2^w$,
where $x\ll y=x\cdot 2^y$),
right shift
($(x,y)\mapsto x\gg y=\Tfloor{{x/{2^y}}}$),
and bitwise Boolean operations
($\textsc{and}$, $\textsc{or}$ and $\textsc{xor}$
(exclusive or)).
We also assume a constant-time operation to
load an integer that deviates from $\sqrt{w}$
by at most a constant factor---this enables the
proof of Lemma~\ref{lem:log}.
We always use ``$\log$'' to denote the binary
logarithm function $\log_2$; when the base is
particularly important, it may still be
indicated explicitly.

The best previous choice dictionary~\cite[Theorem~7.6]{HagK16}
can be initialized
with a universe size $n$ and a second parameter $t\in\TbbbN$
in constant time
and subsequently executes all
operations in $O(t)$ time and occupies
$n+O(n({t/w})^t+\log n)$ bits.
Let us call a choice dictionary \emph{atomic}
if it executes all operations including the
initialization in constant time.
Then, for every constant $t\in\TbbbN$,
the result just cited implies the existence of
an atomic choice dictionary that occupies
$n+O(n/{w^t}+\log n)$ bits when
initialized for universe size~$n$.
In Section~\ref{sec:uncolored}
 we describe an externally sized atomic
choice dictionary that needs just $n+1$ bits,
which is optimal if $w=o(n)$.
The optimality of the bound of $n+1$ bits
follows from a
simple argument of~\cite{HagK17,KatG17}:
Because the client set $S$ of a choice dictionary
with universe size $n$ can be in $2^n$ different
states, any two of which can be distinguished
via calls of $\Tvn{contains}$, if the choice dictionary
uses only $n$ bits it must represent each
possible state of $S$ through a unique bit pattern.
Since $S$ is in one particular state immediately after
the initialization, the latter must force each
of $n$ bits to a specific value, which takes
$\Omega({n/w})$ time.

The notion of iterating over the client set $S$
in order to process its elements one by one
was formalized in~\cite{HagK16} as a virtual
operation \Tvn{iterate} that is a shorthand for
three concrete operations:
$\Tvn{iterate}.\Tvn{init}$, which prepares
for a new iteration over $S$,
$\Tvn{iterate}.\Tvn{next}$, which yields the
next element $\ell$ of $S$ (we say that $\ell$
is \emph{enumerated}; if all elements have already
been enumerated, 0 is returned), and
$\Tvn{iterate}.\Tvn{more}$, which returns~1
if one or more elements of $S$ remain to
be enumerated, and 0 otherwise.
When stating that a choice dictionary allows
iteration in a certain time~$t$, what we mean
is that each of the operations
$\Tvn{iterate}.\Tvn{init}$,
$\Tvn{iterate}.\Tvn{next}$ and
$\Tvn{iterate}.\Tvn{more}$ runs in time bounded by~$t$.
The order in which the elements of $S$ are
enumerated may be chosen arbitrarily by
the data structure.
In applications it is sometimes important to be
able to allow changes to~$S$
while it is being iterated over.
The main choice dictionaries of~\cite{HagK16}
provide \emph{robust} iteration:
Every integer that is a member of $S$ during the
whole iteration is enumerated, while no integer
is enumerated more than once or
at a time when it does not belong to~$S$.
Robust iteration is an ideal that we do not know
how to attain for the new and very
space-efficient choice dictionaries
presented here.
In Section~\ref{sec:iteration}, however, we design
a weaker form of iteration that is still useful:
If the only changes to $S$ during an iteration
are deletions, the iteration is robust and constant-time.
If the only changes are insertions, the iteration
is robust, except that an integer may be enumerated twice,
and the iteration is constant-time
only in an amortized sense.

A generalization of the choice dictionary
called a \emph{colored choice dictionary}, rather
than maintaining a single subset of $U=\{1,\ldots,n\}$,
maintains a \emph{semipartition} $(S_0,\ldots,S_{c-1})$
of $U$, i.e., a sequence of
(possibly empty) disjoint subsets of $U$
whose union is~$U$, called its \emph{client vector}.
Viewing the elements of $S_j$ as having \emph{color} $j$,
for $j=0,\ldots,c-1$, we speak of a
\emph{$c$-color} choice dictionary or a
choice dictionary \emph{for} $c$ colors.
The number $c$ of colors is fixed, together with
the universe size~$n$, during the initialization of an
instance of the data structure, and we now take
``externally sized'' to mean that both $n$ and~$c$
are available without
being stored in the instance.
For emphasis, the original choice dictionary
may be characterized as \emph{uncolored}.
Its operations \Tvn{insert}, \Tvn{delete}
and \Tvn{contains} are replaced by

\begin{description}
\item[\normalfont$\Tvn{setcolor}(j,\ell)$]
($j\in\{0,\ldots,c-1\}$ and $\ell\in U$):
Changes the color of $\ell$ to $j$, i.e.,
moves $\ell$ to $S_j$ (if it is not already there).
\item[\normalfont$\Tvn{color}(\ell)$]
($\ell\in U$):
Returns the color of~$\ell$, i.e.,
the unique $j\in\{0,\ldots,c-1\}$ with $\ell\in S_j$.
\end{description}

Moreover, the operations \Tvn{choice} and
\Tvn{iterate} (with its three suboperations)
take an additional (first) argument $j\in\{0,\ldots,c-1\}$
that indicates the set $S_j$ to which the operations
are to apply; e.g., $\Tvn{choice}(j)$ returns an
arbitrary element of $S_j$ (0 if $S_j=\emptyset$).
Initially, all elements of $U$ belong to~$S_0$.

Sections \ref{sec:colored2} and~\ref{sec:colored}
describe new externally sized
$c$-color choice dictionaries.
Provided that $c$ is a constant---to date the
most useful case in applications---the new
choice dictionaries
are atomic
and occupy $n\log_2\!c+1$ bits
if $c$ is a power of~2 (again,
this is optimal), and at most $n\log_2\!c+n^\epsilon$
bits in general for arbitrary fixed $\epsilon>0$.
As an alternative, still with constant-time
initialization, we can implement
\Tvn{setcolor}, \Tvn{color} and \Tvn{choice}
in $O(\log\log n)$ time using
$n\log_2\! c+O((\log n)^2+1)$ bits.
Except as concerns iteration, the 2-color
choice dictionary subsumes the uncolored choice
dictionary of Section~\ref{sec:uncolored}.
We still provide a self-contained description of the
latter both because it is particularly simple and
may be suited for practical use and classroom teaching
and because all of the techniques developed for the
uncolored case are used again, now in a more complex
setting, in the colored case.

For the colored choice dictionaries,
it is easy to support constant-time iteration
over a set $S_j$ in the client vector
for the \emph{static} case, i.e., when no \Tvn{setcolor}
operations are executed on the choice dictionary
during the iteration.
For the \emph{dynamic} case
we can provide a weak form of iteration over $S_j$
that enumerates only integers
that belong to $S_j$ when they are enumerated and
that enumerates all integers present in~$S_j$
during the whole iteration,
but that may enumerate an integer
repeatedly and for which we can bound the total
iteration time only by $O(m+k\log n)$, where $m$ is
the number of elements present in $S_j$ at the start of the
iteration and $k$ is the number of calls of
\Tvn{setcolor} executed during the iteration
(this bound does not by itself ensure that the iteration
will terminate---we may have $k=\infty$).
Iteration over a colored choice dictionary
is relevant to an algorithm
of~\cite{HagK16} for a problem known loosely as
breadth-first search (BFS) and more precisely as
the computation of a shortest-path spanning
forest of a directed or undirected graph $G$
consistent with a given vertex ordering.
If $G$ has $n$ vertices and $m$ edges, the algorithm
needs $O(n+m)$ time in addition to the time
needed to execute $O(n+m)$ operations on a
3-color dictionary with universe size~$n$
that supports iteration.
Plugging in the new colored choice dictionaries,
we can solve the problem in
$O(m+n\log n)$ time with at most
$n\log_2\! 3+n^\epsilon$ bits,
for arbitrary fixed $\epsilon>0$, or in
$O(n\log n+m\log\log n)$ time with
$n\log_2\! 3+O((\log n)^2+1)$ bits.
The best previous algorithm
\cite{HagK16}[Theorem~8.5] is faster, $O(n+m)$ time, but
needs more space, $n\log_2\!3+O({n/{(\log n)^t}})$ bits
for arbitrary fixed $t\in\TbbbN$.

Our new results were obtained by combining
techniques of
Katoh and Goto~\cite{KatG17} and
Hagerup and Kammer~\cite{HagK16}
with new ideas.
Katoh and Goto used a new so-called
\emph{in-place chain technique} to obtain improved
\emph{initializable arrays},
arrays with an additional operation to
store the same given value in every cell.
A connection between choice dictionaries and
initializable arrays was first noted by
Hagerup and Kammer~\cite{HagK17}, who observed
that the \emph{light-path technique}, invented in~\cite{HagK16}
in the context of choice dictionaries, also yields
initializable arrays better than those known at the time.
Here we use the in-place chain technique,
slightly modified and extended with new operations,
to derive the uncolored choice
dictionary of Section~\ref{sec:uncolored}.
Parts of the present paper that draw heavily on
techniques and results of~\cite{HagK16} include
Subsections \ref{subsec:base} and~\ref{subsec:small}
and there, in particular, a method of storing
external information in regularly spaced free bits
in the representation of a choice dictionary
during times when its
client vector is \emph{deficient},
i.e., contains one or more empty sets.
A difference is that whereas isolated free bits
are sufficient in~\cite{HagK16}, here a new step
had to be introduced that aggregates free bits
into groups of $\Omega(\log n)$ contiguous bits.

Our space bounds for a data structure apply at times
when the data structure is in a \emph{quiescent} state,
i.e., between the execution of operations.
During the execution of an operation, the data
structure may temporarily need more space---we
speak of \emph{transient} space requirements.
By definition of the word RAM, the transient
space requirements are always at least $\Theta(w)$ bits,
and we will mention them only if they exceed
this bound.
Similarly, if no space bound is indicated for
an algorithm, it gets by with $O(w)$ bits
in addition to the space needed for its
input and output.

The present text is an expanded version of a
preliminary report~\cite{Hag17} and essentially
repeats the material of the latter
for uncolored choice dictionaries
as the main part of Section~\ref{sec:uncolored}.
A different generalization of the approach
of \cite{Hag17} to $c\ge 2$ colors was found recently
by Kammer and Sajenko~\cite{KamS18},
but only for the case in which $c$ is a power of~2.
The colored choice dictionaries of \cite{KamS18}
are simpler than ours and allow a slightly
more general form of iteration.
This paper, on the other hand,
allows general values of~$c$
and offers faster operations
for nonconstant~$c$, especially if $w=\omega(\log n)$,
and better support for
iteration in the uncolored case.

\section{An Uncolored Choice Dictionary}
\label{sec:uncolored}

This section describes the very simple new
atomic uncolored choice dictionary and provides
fairly detailed pseudo-code for its operations.
The addition
of support for iteration
to the data structure
is postponed to Section~\ref{sec:iteration}.

\begin{theorem}
\label{thm:uncolored}
There is an externally sized atomic
(uncolored) choice dictionary that,
when initialized for universe size~$n$,
occupies $n+1$ bits.
\end{theorem}

Throughout the paper we make use of
the natural bijections, for each
given $n\in\TbbbN$,
that relate a subset $S$ of $\{0,\ldots,n-1\}$,
the integer $\sum_{\ell\in S}2^\ell$
in $\{0,\ldots,2^n-1\}$, and the sequence
$(b_0,\ldots,b_{n-1})$ of $n$ bits with
$b_\ell=1\Leftrightarrow\ell\in S$,
for $\ell=0,\ldots,n-1$.
When speaking about the finite subset
of $\TbbbN_0=\TbbbN\cup\{0\}$, the nonnegative
integer or the bit sequence
corresponding to an object $X$
of one of the two other kinds
we mean the object obtained from $X$ by
an application of the relevant bijection.
Already the definition of the word RAM makes
use of this correspondence by viewing the
contents of memory words as integers when
arithmetic operations are applied to them
and as bit sequences when the operations
\textsc{and}, \textsc{or} and \textsc{xor} are used.
Similarly, we may view a sequence of
bits stored in memory as representing a nonnegative
integer (given via its binary representation)
or a finite subset of $\TbbbN_0$
(given via its bit-vector representation).
When the inverse bijections are used to derive
a bit sequence from a finite subset of $\TbbbN_0$ or
a nonnegative integer, the length of the
bit sequence must be supplied either explicitly
or by context, since a bit sequence can always be
extended by additional zeros without any change
to the set or the integer that it represents;
it will usually be clear that
a bit sequence of a particular length is called for.
For given integers $a$ and $n\in\TbbbN$, we
can extend the bijections under consideration
to subsets of $\{a,\ldots,a+n-1\}$ by mapping
$S\subseteq\{a,\ldots,a+n-1\}$ to
$\{\ell-a\mid\ell\in S\}\subseteq\{0,\ldots,n-1\}$.
Thus such a set may also be represented via
a bit sequence or a nonnegative integer.

\subsection{A Simple Reduction}
\label{subsec:reduction}

If we represent the client set of a choice
dictionary with universe size $n$ via its
bit-vector representation~$B$ of length~$n$,
the choice-dictionary
operations translate into the reading and writing
of individual bits in $B$ and the operation
\Tvn{choice}, which now returns the position of
a nonzero bit in~$B$ (0 if all bits in $B$
are~\texttt{0}).
It is trivial to carry out all operations other
than the initialization and \Tvn{choice}
in constant time.
In the special case $n=O(w)$, the latter
operations can also be supported in constant time.
This is a consequence of part~(a)
of the following lemma, used with $f=1$.

\begin{lemma}[\cite{HagK16}]
\label{lem:log}
Let $m$ and $f$ be given integers
with $1\le m,f<2^w$ and suppose that a
sequence $(a_1,\ldots,a_m)$ with
$a_i\in\{0,\ldots,2^f-1\}$ for $i=1,\ldots,m$
is given
in the form of the $(m f)$-bit binary representation
of the integer
$x=\sum_{i=0}^{m-1} 2^{i f}a_{i+1}$.
Then the following holds:
\begin{itemize}
\item[(a)]
Let $I_{>0}=\{i\in\TbbbN\mid 1\le i\le m$ and $a_i>0\}$.
Then, in $O(\Tceil{{{m f}/w}})$ time,
we can test whether $I_{>0}=\emptyset$ and, if not,
compute $\min I_{>0}$ and $\max I_{>0}$.
\item[(b)]
Let $I_0=\{i\in\TbbbN\mid 1\le i\le m$ and $a_i=0\}$.
Then, in $O(\Tceil{{{m f}/w}})$ time,
we can test whether $I_0=\emptyset$ and, if not,
compute $\min I_0$.
\item[(c)]
If an additional integer $k\in\{0,\ldots,2^f-1\}$
is given, then $O(\Tceil{{{m f}/w}})$ time
suffices to compute the integer
$\sum_{i=0}^{m-1}2^{i f}b_{i+1}$, where
$b_i=1$ if $k\ge a_i$ and $b_i=0$ otherwise
for $i=1,\ldots,m$.
\end{itemize}
\end{lemma}

We use the
externally sized atomic choice dictionary for universe
sizes of $O(w)$ implied by these considerations
to handle the few bits left over when we divide
a bit-vector representation of $n$ bits into
pieces of a fixed size.
The details are as follows:

Let $b$ be a positive integer that can be computed
from $w$ and $n$ in constant time using $O(w)$ bits
(and therefore need not be stored) and that
satisfies $b\ge\log_2\! n$, but $b=O(w)$.
In order to realize an
externally sized choice dictionary $D$ with
universe size~$n$ and client set~$S$, partition the
bit-vector representation $B$ of $S$ into
$N=\Tfloor{n/{(2 b)}}$
segments $B_1,\ldots,B_N$
of exactly $2 b$ bits each, with
$n'=n\bmod(2 b)$ bits left over.
If $n'\not=0$, maintain (the set
corresponding to) the last $n'$
bits of $B$ in an externally sized
atomic choice dictionary $D_2$
realized as discussed above.
Assume without loss of generality that $N\ge 1$.
The following lemma is proved in the
remainder of this section:

\begin{lemma}
\label{lem:main}
There is a data structure that,
if given access to $b$ and $N$,
can be initialized
in constant time and subsequently occupies $2 b N+1$ bits and
maintains a sequence
$(a_1,\ldots,a_N)\in\{0,\ldots,2^{2 b}-1\}^N$,
initially $(0,\ldots,0)$, under
the following operations, all of which
execute in constant time:
$\Tvn{read}(k)$ ($k\in\{1,\ldots,N\}$), which
returns $a_k$;
$\Tvn{write}(k,x)$ ($k\in\{1,\ldots,N\}$ and
$x\in\{0,\ldots,2^{2 b}-1\}$), which 
sets $a_k$ to $x$;
and \Tvn{nonzero},
which returns a $k\in\{1,\ldots,N\}$
with $a_k\not=0$
if there is such a $k$, and 0 otherwise.
\end{lemma}

For $k=1,\ldots,N$, view $B_k$ as the binary
representation of an integer and maintain
that integer as~$a_k$ in an instance of
the data structure of Lemma~\ref{lem:main}.
This yields an externally sized
atomic choice dictionary
$D_1$ for (the set corresponding to)
the first $2 b N$ bits of $B$:
To carry out \Tvn{insert}, \Tvn{delete}
or \Tvn{contains}, update or inspect the
relevant bit in one of $a_1,\ldots,a_N$,
and to execute \Tvn{choice}, call \Tvn{nonzero}
and, if the return value $k$ is positive,
apply an algorithm of Lemma~\ref{lem:log}(a) to~$a_k$
and add $2 b(k-1)$.
It is obvious how to realize the full
choice dictionary $D$ through a combination
of $D_1$ and $D_2$.
The only nontrivial case is that of
the operation \Tvn{choice}.
To execute \Tvn{choice} in $D$, first call
\Tvn{choice} in $D_1$ (say).
If the return value is positive, it is a suitable
return value for the parent call.
Otherwise call \Tvn{choice} in $D_2$,
increase the return value by $2 b N$ if it is positive,
and return the resulting integer.
$D$ is atomic because $D_1$ and $D_2$ are, and
the total number of bits used by $D$
is $2 b N+1+n'=n+1$.
Theorem~\ref{thm:uncolored} follows.

\subsection{The Storage Scheme}
\label{subsec:scheme}

To prove Lemma~\ref{lem:main}, we first show a
slightly weaker form of the lemma in which the
space bound is relaxed to allow
$2 b N+w$ bits instead of $2 b N+1$ bits.
This subsection describes how the sequence
$(a_1,\ldots,a_N)$ is represented in memory in
$2 b N+w$ bits.
Most of the available memory stores an array $A$
of $N$ cells $A[1],\ldots,A[N]$ of $2 b$ bits each.
In addition, a $w$-bit word is used to hold
an integer $\mu\in\{0,\ldots,N\}$ best thought of as a
``barrier'' that divides $V=\{1,\ldots,N\}$
into a part to the left of the barrier, $\{1,\ldots,\mu\}$,
and a part to its right, $\{\mu+1,\ldots,N\}$.
We often consider a $(2 b)$-bit quantity
$x$ to consist of a \emph{lower half}, denoted by
$\underline{x}$ and composed of the $b$
least significant bits of $x$
(i.e., $\underline{x}=x\,\mathbin{\textsc{and}}\,(2^b-1)$),
and an \emph{upper half},
$\overline{x}=x\gg b$, and we may
write $x=(\underline{x},\overline{x})$.
A central idea,
due to Katoh and Goto~\cite{KatG17},
is that the upper halves of
$A[1],\ldots,A[N]$ are used to implement a
matching on $V$
according to the following convention:
Elements $k$ and $\ell$ of $V$ are matched exactly if
$\overline{A[k]}=\ell$, $\overline{A[\ell]}=k$,
and precisely one of $k$ and $\ell$ lies to the
left of the barrier, i.e., $k\le \mu<\ell$
or $\ell\le \mu<k$.
In this case we call $\ell$ the \emph{mate} of
$k$ and vice versa.
The assumption $b\ge\log_2\! n$ ensures that
the upper half of each cell in~$A$ can hold an arbitrary
element of~$V$.
A function that inputs an element $k$ of $V$
and returns the mate of $k$ if $k$ is matched
and $k$ itself if not is easily coded as follows:

\begin{tabbing}
\quad\=\quad\=\quad\=\quad\=\kill
\Tvn{mate}$(k)$:\\
\>$k':=\overline{A[k]}$;\\
\>\textbf{if} $(1\le k\le \mu<k'\le N$ or
$1\le k'\le \mu<k\le N)$ and $\overline{A[k']}=k$
\textbf{then return} $k'$;\\
\>\textbf{return} $k$;
\end{tabbing}

\noindent
For all $k\in V$, call $k$ \emph{strong}
if $k$ is matched and $k\le \mu$ or
$k$ is unmatched and $k>\mu$,
and call $k$ \emph{weak} if it is not strong.
The integers $A[1],\ldots,A[N]$ and $\mu$
represent the sequence $(a_1,\ldots,a_N)$
according to the following storage invariant:
For all $k\in V$,

\begin{itemize}
\item
$a_k=0$ exactly if $k$ is weak;
\item
if $k$ is strong and $k>\mu$, then $a_k=A[k]$;
\item
if $k$ is strong and $k\le \mu$, then
$a_k=(\underline{A[k]},\underline{A[\Tvn{mate}(k)]})$.
\end{itemize}

\begin{figure}
\begin{center}
\epsffile{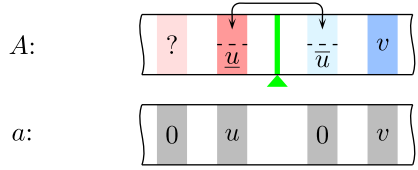}
\end{center}
\caption{The storage scheme.
Above: The array $A$.
Below: The sequence $a$ represented by $A$.}
\label{fig:1}
\end{figure}

The storage invariant is illustrated in Fig.~\ref{fig:1}.
The following drawing conventions are used here
and in subsequent figures:
The barrier is shown as a thick vertical line segment
with a triangular base.
Each pair of mates is connected with a double arrow,
and a cell $A[k]$ of $A$ is shown in a
darker hue if $k$ is strong.
A question mark indicates an entry that can be
completely arbitrary, except that it may not give
rise to a matching edge, and the upper and lower halves
of some cells of $A$ are shown separated by
a dashed line segment.

\subsection{The Easy Operations}

The data structure is initialized by setting $\mu=N$,
i.e., by placing the barrier at the right end.
Then the matching is empty, and all elements of $V$
are to the left of the barrier and weak.
Thus the initial value of $(a_1,\ldots,a_N)$
is $(0,\ldots,0)$, as required.
The implementation of $\Tvn{read}$
closely reflects
the storage invariant:

\begin{tabbing}
\quad\=\quad\=\quad\=\quad\=\kill
\Tvn{read}$(k)$:\\
\>\textbf{if} $\Tvn{mate}(k)\le \mu$ \textbf{then return} 0;
 $(*$ $k$ is weak exactly if $\Tvn{mate}(k)\le \mu$ $*)$\\
\>\textbf{if} $k>\mu$ \textbf{then return}
$A[k]$; \textbf{else return}
$(\underline{A[k]},\underline{A[\Tvn{mate}(k)]})$;
\end{tabbing}

\noindent
The code for \Tvn{nonzero} is short but a
little tricky:

\begin{tabbing}
\quad\=\quad\=\quad\=\quad\=\kill
\Tvn{nonzero}:\\
\>\textbf{if} $\mu=N$ \textbf{then return} 0;
\textbf{else return} $\Tvn{mate}(N)$;
\end{tabbing}

The implementation of $\Tvn{write}(k,x)$
is easy if $k$ is weak and
$x=0$ (then nothing needs to be done) or
$k$ is strong and $x\not=0$.
In the latter case the procedure \Tvn{simple\_write}
shown below can be used.
The only point worth noting is that writing
to $A[k]$ when $k$ is strong and $k>\mu$
may create a spurious matching edge that
must be eliminated.

\begin{tabbing}
\quad\=\quad\=\quad\=\quad\=\kill
\Tvn{simple\_write}$(k,x)$:\\
\>\textbf{if} $k\le \mu$ \textbf{then}
$(\underline{A[k]},
\underline{A[\Tvn{mate}(k)]}):=
 (\underline{x},\overline{x})$;\\
\>\textbf{else}\\
\>\>$A[k]:=x$;\\
\>\>$k':=\Tvn{mate}(k)$;\\
\>\>\textbf{if} $k'\not=k$ \textbf{then} $\overline{A[k']}:=k'$;
 $(*$ eliminate a spurious matching edge $*)$
\end{tabbing}

\subsection{Insertion and Deletion}
\label{subsec:insdel}

The remaining, more complicated, operations of
the form $\Tvn{write}(k,x)$ are those in which
$a_k$ is changed from zero to
a nonzero value---call such an operation an
\emph{insertion}---or vice versa---a \emph{deletion}.
When the data structure under development is used
to realize a choice dictionary, insertions and
deletions
are triggered by (certain)
insertions and
deletions, respectively, executed on the choice dictionary.
Insertions and deletions
are the operations that change the
barrier and usually the matching.
In fact, $\mu$ is decreased by~1 in every
insertion and increased by~1 in every deletion,
so $\mu$ is always the number of
$k\in V$ with $a_k=0$.

\begin{figure}
\begin{center}
\epsffile{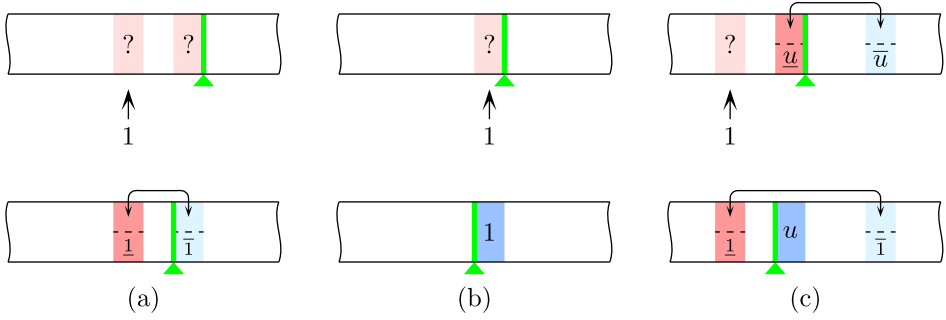}
\caption{Insertion to the left of the barrier.}
\label{fig:2}
\end{center}
\end{figure}

\begin{figure}
\begin{center}
\epsffile{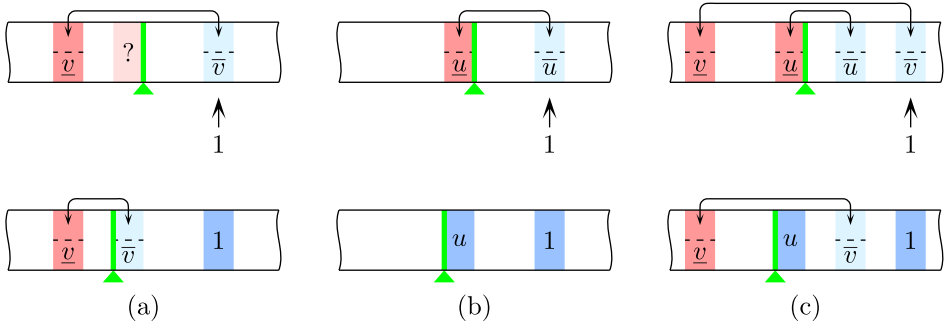}
\caption{Insertion to the right of the barrier.}
\label{fig:3}
\end{center}
\end{figure}

\begin{figure}
\begin{center}
\epsffile{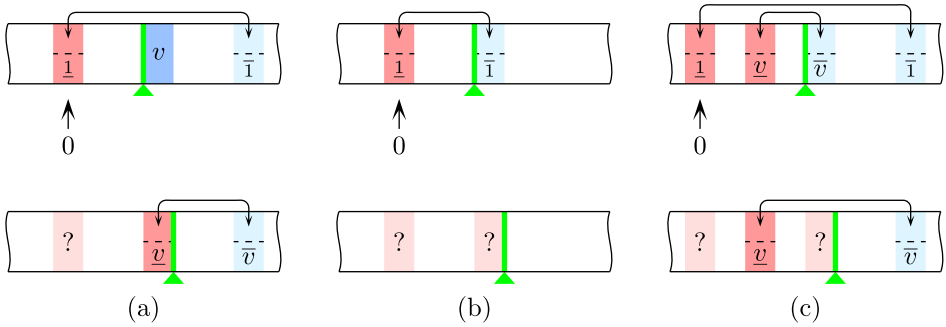}
\caption{Deletion to the left of the barrier.}
\label{fig:4}
\end{center}
\end{figure}

\begin{figure}
\begin{center}
\epsffile{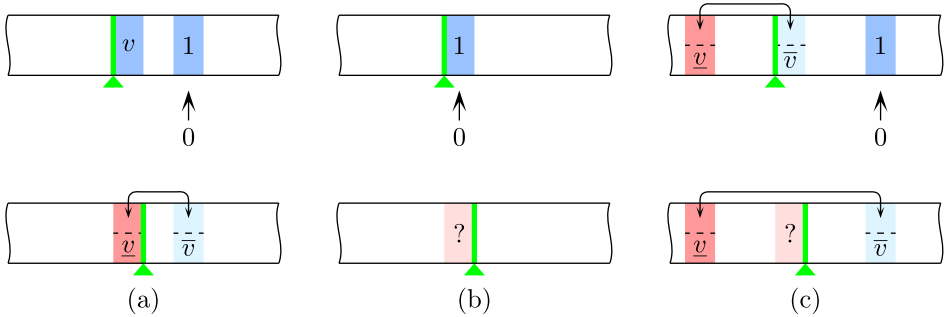}
\caption{Deletion to the right of the barrier.}
\label{fig:5}
\end{center}
\end{figure}

The various different forms that an insertion
may take are illustrated in Figs.\
\ref{fig:2} and~\ref{fig:3}.
The situation before the insertion is always
shown above the situation after the insertion.
A ``1'' outside of the ``stripes'' indicates
the position of an insertion and symbolizes
the nonzero value to be written, while a ``1'' inside
the stripes symbolizes that value after
it has been written.
The various forms of a deletion are illustrated
in Figs.\ \ref{fig:4} and~\ref{fig:5}.
Here a ``0'' indicates the position of a deletion,
while a ``1'' symbolizes the nonzero value that
is to be replaced by zero.

There are many somewhat different
cases, but for each it is easy to
see that the storage invariant
is preserved and that
the sequence $(a_1,\ldots,a_N)$
changes as required.
It is also easy to turn the figures into a
\Tvn{write} procedure that branches
into as many cases.
Here we propose the
following realization of \Tvn{write}
that is terser, but needs a more
careful justification.

\begin{tabbing}
\quad\=\quad\=\quad\=\quad\=\kill
\Tvn{write}$(k,x)$:\\
\>$x_0:=\Tvn{read}(k)$; $(*$ the value to be replaced by $x$ $*)$\\
\>$k':=\Tvn{mate}(k)$;\\
\>\texttt{if} $x\not=0$ \texttt{then}\\
\>\>\texttt{if} $x_0=0$ \texttt{then} $(*$ an insertion $*)$\\
\>\>\>$\mu':=\Tvn{mate}(\mu)$; $(*$ $\wik=\mu$ will cross the barrier $*)$\\
\>\>\>$u:=\Tvn{read}(\mu)$; $(*$ save $a_{\wik}$ $*)$ \\
\>\>\>$\mu:=\mu-1$; $(*$ move the barrier left $*)$\\
\>\>\>\Tvn{simple\_write}$(\mu+1,u)$;
 $(*$ reestablish the value of $a_{\wik}$ $*)$\\
\>\>\>\textbf{if} $k\not=\mu'$ \texttt{then} $\{$
$\overline{A[k']}:=\mu'$; $\overline{A[\mu']}:=k'$;
 $\underline{A[\mu']}:=\underline{A[k]}$; $\}$
 $(*$ match $k'$ and $\mu'$ $*)$\\
\>\>\Tvn{simple\_write}$(k,x)$; $(*$ $k$ was or has been made strong $*)$\\
\>\textbf{else} $(*$ $x=0$ $*)$\\
\>\>\texttt{if} $x_0\not=0$ \texttt{then} $(*$ a deletion $*)$\\
\>\>\>$\mu':=\Tvn{mate}(\mu+1)$; $(*$ $\wik=\mu+1$ will cross the barrier $*)$\\
\>\>\>$v:=\Tvn{read}(\mu')$; $(*$ save $a_{\mu'}$ $*)$ \\
\>\>\>$\mu:=\mu+1$; $(*$ move the barrier right $*)$\\
\>\>\>$\overline{A[k']}:=\mu'$; $\overline{A[\mu']}:=k'$;
 $(*$ match $k'$ and $\mu'$ $*)$\\
\>\>\>\texttt{if} $\mu'\not=k$ \texttt{then}
 $\Tvn{simple\_write}(\mu',v)$; $(*$ reestablish the
 value of $a_{\mu'}$ $*)$
\end{tabbing}

To see the correctness of the procedure
\Tvn{write} given above,
consider a call $\Tvn{write}(k,x)$ and assume that
it gives rise to an insertion or a deletion,
since in the remaining cases the procedure is
easily seen to perform correctly.
Let $\mu_0$ be the value of $\mu$ (immediately)
before the call.
Because the call changes the value of $\mu$ by~1,
a single element $\wik$ of $V$
\emph{crosses} the barrier, i.e., is to the left
of the barrier before or after the call, but not both.
In the case of an insertion, $\wik=\mu_0$;
in that of a deletion, $\wik=\mu_0+1$.

Assume that $k$ does not cross the barrier,
i.e., that $k\not=\wik$.
Because the call changes $a_k$ from zero to a
nonzero value or vice versa, $k$ must change its
matching status, i.e., be matched before or after
the call, but not both.
In detail, if $k$ is matched before the call,
its mate at that time, if different from $\wik$,
must find a new mate, which automatically
leaves $k$ unmatched.
If $k$ is unmatched before the call, $k$ itself
must find a mate.
We can unify the two cases by saying that if
$k'=\Tvn{mate}(k)$ (evaluated before the call
under consideration has changed $\mu$ and~$A$)
is not $\wik$, then $k'$ must find a (new) mate.
If $k'\not=\wik$, moreover, $k'$ is to the
left of the barrier in the case of an insertion
and to the right of it in the case of a deletion.

Assume now that the call does not change
$a_{\wik}$, i.e., that $\wik\not=k$.
Then, because $\wik$ crosses the barrier,
it must also change its matching status:
If $\wik$ is matched before the call,
its mate at that time, if different from $k$,
must find a new mate, and otherwise $\wik$
itself must find a mate.
As above, this can be expressed by saying that
if $\mu'=\Tvn{mate}(\wik)$ (evaluated before the call
has changed $\mu$ and~$A$)
is not $k$, then $\mu'$ must find a (new) mate.
Moreover, after the call $\mu'$ is to
the right of the barrier in the case of an
insertion and to the left of it in the case of a deletion.

Exclude the special cases identified above
by assuming that
$\{k,k'\}\cap\{\wik,\mu'\}=\emptyset$.
Then it can
be seen that all required changes to the matching
can be effectuated by matching $k'$ and $\mu'$,
which is what the procedure \Tvn{write} does.
In the case of an insertion, this makes $k$ strong,
which implies that $a_k$ can be set to $x$ simply
by executing $\Tvn{simple\_write}(k,x)$
at the very end.

In addition, with $\ell=\min\{k',\mu'\}$, it
must be ensured that the call does not change~$a_\ell$
except if $\ell=k$.
In the case of an insertion, $\ell=k'$, and
if $k'\not=k$, the mate of $k'$ switches from being $k$
to being $\mu'$, so that it suffices to
execute $\underline{A[\mu']}:=\underline{A[k]}$,
which happens in the procedure.
The same assignment is executed if $k'=k$,
in which case it is useless but harmless.
In the case of a deletion, $\ell=\mu'$.
Here the procedure plays it safe by remembering
the value of $a_{\mu'}$ before the call in a variable~$v$
and restoring $a_{\mu'}$ to that value
at the end, unless $\mu'=k$, via the call
$\Tvn{simple\_write}(\mu',v)$.
This is convenient because $a_{\mu'}$
is not stored in a unique way before the call.

At this point
$k$, $k'$ and $\mu'$ have been ``taken care of'',
but $\wik$ still needs attention.
In the case of a deletion, either $\wik=\mu'$ or
$\wik$ is weak, so nothing more needs to be done.
In an insertion, the procedure saves the original
value of $a_{\wik}$ in $u$ and restores it afterwards
through the statement $\Tvn{simple\_write}(\mu+1,u)$.
This is necessary and meaningful only if $\wik$ is strong.
If $\wik$ is weak, however, the effect of the
statement---except for the harmless
possible elimination
of a spurious matching edge---is canceled through
the subsequent assignment to
$\overline{A[\mu']}$ and $\underline{A[\mu']}$.

We still need to consider the special cases that
were ignored above, namely calls with
$\{k,k'\}\cap\{\wik,\mu'\}\not=\emptyset$.
These form part~(b)
of Figs. \ref{fig:2}--\ref{fig:5}.
In fact, the number of special cases is quite limited.
If $\wik$ is weak before an insertion
or strong before a deletion, it is unmatched.
Thus if $k=\wik$, we have $k=k'=\wik=\mu'$,
and $k'=\mu'$ implies $k=\wik$.
On the other hand, each of the statements
$k=\mu'$ and $k'=\wik$ implies the other one.
Thus there are two cases to consider:
(1) $k=k'=\wik=\mu'$
and (2) $k=\mu'\not=\wik=k'$.

In case~(1), all writing to $A$ happens
to $A[k]$.
For insertion (Fig.\ \ref{fig:2}(b)), the execution of
$\Tvn{simple\_write}(k,x)$ at the very end
ensures the correctness of the call.
For deletion (Fig.\ \ref{fig:5}(b)), the execution of
$\overline{A[\mu']}:=k'$ at the end ensures that
$k$ is unmatched, which is all that is required.
In case (2), after an insertion (Fig.\ \ref{fig:3}(b)),
$k$ and $\wik$ are both to the right of
the barrier, $a_k$ and $a_{\wik}$ are both nonzero,
and the execution of
$\Tvn{simple\_write}(k,x)$ and
$\Tvn{simple\_write}(\mu+1,u)$
ensures that $A[k]$ and $A[\wik]$ have the correct
values after the call.
After a deletion (Fig.\ \ref{fig:4}(b)), $k$ and $\wik$ are both
to the left of the barrier and $a_k=a_{\wik}=0$,
and the execution of
$\overline{A[k']}:=\mu'$ and $\overline{A[\mu']}:=k'$
in fact ensures that $k$ and $\wik$ are both unmatched,
which is all that is required.

Since all operations of the data structure have been
formulated as pieces of code without loops
and $b=O(w)$, 
it is clear that the operations execute in constant time.

\subsection{Reducing the Space Requirements}
\label{subsec:hiding}

The space requirements of the data
structure of Subsections \ref{subsec:scheme}--\ref{subsec:insdel}
can be reduced
from $2 b N+w$ bits to $2 b N+1$
bits, as promised in
Lemma~\ref{lem:main},
by a method of~\cite{HagK17,KatG17}.
First, $b$ is chosen to satisfy
not only $b\ge\log_2\! n$, but $b\ge 2\log_2\! n$,
which is clearly still compatible with $b=O(w)$.
As a result, for each $k\in V$ to
the left of the barrier, $A[k]$ has at least
$2\log_2\! n-\Tceil{\log_2\! N}\ge\Tceil{\log_2\! N}$ unused bits.
If $\mu\ge 1$, we store $\mu$ in the unused bits of~$A[1]$
(the unused bits of $A[2],\ldots,A[\mu]$ continue to be unused).
When $\mu=0$, even $A[1]$ is to the right of the barrier
and there are no unused bits in $A$, so we use a
single bit outside of $A$ to indicate whether $\mu$ is nonzero.
The resulting data structure occupies
exactly $2 b N+1$ bits.

\subsection{The Choice of $b$}

A practical choice dictionary
based on the ideas of this section
is likely to
content itself with the main construction of
Subsections \ref{subsec:scheme}--\ref{subsec:insdel}
and refrain from
applying the method of
Subsection~\ref{subsec:hiding} to squeeze
out the last few bits.
Then there is no reason to choose $b$ larger
than $w$, and $b=w$ seems the best choice.
This yields a self-contained atomic
choice dictionary that occupies $n+2 w$ bits
when initialized for universe size~$n$.

If $w$ is even and $w\ge 2\log_2\! n$,
another plausible choice is $b={w/2}$,
which allows an entry in the array $A$
to be manipulated with a single instruction
and simplifies the access to cells of~$A$.
It seems, however, that the gains in
certain scenarios from choosing $b={w/2}$
instead of $b=w$ are small and can be reduced
still further through an optimization of
the case $b=w$ that
omits superfluous operations on upper or lower
halves of cells in $A$.

If the space needed for an externally sized
choice dictionary is to be reduced all the
way to $n+1$ bits for universe size~$n$,
$b=2 w$ seems the best choice.

\subsection{A Self-Contained Choice Dictionary}

In order to convert the externally sized atomic
choice dictionary of Theorem~\ref{thm:uncolored}
to a self-contained one,
we must augment the data structure with
an indication of the universe size~$n$.
This can clearly always be done with
$w$ additional bits.
If a space bound is desired that depends only on~$n$,
$n$ must be stored as a so-called
self-delimiting numeric value.
Assume first that the most significant bits in
a word are considered to be its ``first'' bits,
i.e., the ones to be occupied by a data structure
of fewer than~$w$ bits
(the ``big-endian'' convention).
Then one possibility is to use the
code $\gamma'$
of Elias~\cite{Eli75}:
With $\Tvn{bin}(n)$ denoting the usual binary
representation of $n\in\TbbbN$
(e.g., $\Tvn{bin}(13)=\texttt{1101}$),
store $n$ in the form of the string
$\texttt{0}^{|\mathit{bin}(n)|-1}\Tvn{bin}(n)$,
which can be decoded in constant time
with an algorithm of Lemma~\ref{lem:log}(a).
Since $|\Tvn{bin}(n)|=\Tceil{\log(n+1)}$,
this yields a space bound for the
self-contained choice dictionary
of $n+2\Tceil{\log(n+1)}$ bits.
If instead the least significant bits of a word are
considered to be its first bits
(the ``little-endian'' convention),
the scheme needs to be changed slightly:
The string $\texttt{0}^{|\mathit{bin}(n)|-1}\Tvn{bin}(n)$
is replaced by
$\widehat{\Tvn{bin}}(n)\texttt{0}^{|\mathit{bin}(n)|-1}$,
where $\widehat{\Tvn{bin}}(n)$ is the same as
$\Tvn{bin}(n)$, except that the leading \texttt{1}
is moved to the end.

Incidentally, if an application can guarantee
that $\mu$ never becomes zero, the method of
Subsection~\ref{subsec:hiding}
can be used to ``hide'' $n$
as well as $\mu$ in the array $A$
if we choose $b\ge 4\Tceil{\log(n+1)}$.
This yields a restricted self-contained
atomic choice dictionary that
occupies $n+1$ bits.
The restriction is satisfied, e.g., if
the universe $\{1,\ldots,n\}$ always contains
$4 b-1$ consecutive elements
that do not belong to the client set.

\subsection{Making the Choice Dictionary Dynamic}

It is easy to extend the choice dictionary of
Theorem~\ref{thm:uncolored} to allow gradual
changes to the universe size, i.e., to support the
following two additional operations, where
$n$ is the universe size and $S$ is the client set:

\begin{description}
\item[\normalfont$\Tvn{expand}(b)$]
($b\in\{\texttt{0},\texttt{1}\}$):
Increases $n$ by~1 and subsequently, if $b=\texttt{1}$,
replaces $S$ by $S\cup\{n\}$.
\item[\normalfont$\Tvn{contract}$]
($n>0$):
Replaces $S$ by $S\setminus\{n\}$ and subsequently
decreases $n$ by~1.
\end{description}

We call the resulting data structure a
\emph{dynamic} externally sized (uncolored)
choice dictionary.
We allow the universe size of a dynamic choice
dictionary to be~0.
When it is, \Tvn{choice} should return~0, and
calls of \Tvn{insert}, \Tvn{delete} and \Tvn{contains}
are illegal.

\begin{theorem}
There is an atomic dynamic externally sized (uncolored)
choice dictionary that occupies $n+1$ bits when
its universe size is~$n$, for all $n\in\TbbbN_0$.
\end{theorem}

\begin{proof}
We use the same construction as for the choice
dictionary of Theorem~\ref{thm:uncolored}, except
that $b$ should now be chosen as a function of $w$ alone.
Apart from changing the externally stored universe
size in the obvious way, the operations \Tvn{expand}
and \Tvn{contract} carry out the steps described
in the following.
First, $\Tvn{expand}(b)$ stores $b$ in the new bit
that becomes available to the data structure.
Unless the call of \Tvn{expand} or \Tvn{contract}
changes $\Tfloor{n/{(2 b)}}$, it need not do
anything else---the number $n'$ of bits in the
trivial choice dictionary $D_2$
of Subsection~\ref{subsec:reduction}
simply increases or decreases by~1 (allow $n'=0$).
Assume now that the
call of \Tvn{expand} or \Tvn{contract}
changes $\Tfloor{n/{(2 b)}}$, so that $n'$ jumps from
$2 b-1$ to 0 or vice versa and the array $A$
acquires a new cell or loses one.
If the call in question is $\Tvn{expand}(\texttt{1})$
and the universe size after the call is~$n$,
let $x$ be the integer formed by the last
$2 b$ of the $n+1$ bits that represent the data
structure and execute $\Tvn{simple\_write}(n,x)$,
which serves exclusively to eliminate a possible
spurious matching edge.
If the call is $\Tvn{expand}(\texttt{0})$,
simulate it by $\Tvn{expand}(\texttt{1})$ followed
by $\Tvn{delete}(n)$.
If the call in question is $\Tvn{contract}$ and the
universe size before the call is $n$,
simply execute $\Tvn{insert}(n)$ before the change
to the universe size.
This implementation of \Tvn{expand} and \Tvn{contract}
works because the rest of the representation of
the data structure, except for the issue of a
spurious matching edge, is independent of the presence
or absence of a last cell in $A$ whose index is strong.
\end{proof}

\section{Power-of-2-Colored Choice Dictionaries}
\label{sec:colored2}

This section describes colored choice dictionaries
that can be used only in the simpler case
in which the number $c$ of
colors is a power of~2.
Let us call such choice dictionaries
\emph{power-of-2-colored}.
The case of general values of~$c$ is considered
in the next section, and
the discussion of
iteration
is again postponed to Section~\ref{sec:iteration}.
To avoid trivialities, we always assume that
the number $c$ of colors is at least~2.

The following lemma, based on the fast
integer-multiplication
algorithm of Sch\"onhage and Strassen~\cite{SchS71},
bounds the complexity of
multiple-precision multiplication and division
on the $w$-bit word RAM.
Except for the space bound, it was observed
in~\cite{Hag15}.

\begin{lemma}
\label{lem:multiple}
For all integers $m$ and $n$ with $1\le m\le n$,
if $x$ and $y$ are given
integers with $0\le x<2^n$ and $0\le y<2^m$,
then $x y$ and, if $y>0$, $x/y$ can be computed in
$O(\Tceil{{{n\log(2+{m/w})}/w}})$ time
with $O(n+m+w)$ bits of working memory.
\end{lemma}

For $m,f\in\TbbbN$,
let $1_{m,f}=\sum_{i=0}^{m-1}2^{i f}=
{{(2^{m f}-1)}/{(2^f-1)}}$.
If the $(m f)$-bit binary representation
of $1_{m,f}$ is divided into $m$ fields of $f$
bits each, each field contains the value~1.
As follows from \cite[Theorem~2.5]{Hag15},
the possibly multiword integer $1_{m,f}$
can be computed in $O(\Tceil{{{m f}/w}})$ time,
and an integer in $\{0,\ldots,2^n-1\}$
can be multiplied by $1_{m,f}$ in
$O(\Tceil{{{(n+m f)}/w}})$ time
with $O(n+m f+w)$ bits of working memory.

\subsection{Changing Base}
\label{subsec:base}

In this subsection we study the following problem,
which plays a central role for all of
our colored dictionaries:
Given integers $c$ and $d$ with $c,d\ge 2$
and an integer $x$ of the form
$x=\sum_{j=0}^{s-1}a_j c^j$,
where $s\in\TbbbN$ and
$a_0,\ldots,a_{s-1}$ are integers with
$0\le a_j<\min\{c,d\}$ for $j=0,\ldots,s-1$,
compute $y=\sum_{j=0}^{s-1}a_j d^j$.
From the perspective of positional numeral systems,
the problem can be viewed as one of changing the base
from $c$ to $d$,
but in the peculiar sense of leaving the digits
unchanged while interpreting them according
to a new base.
Alternatively, the problem can be seen as
the evaluation of a polynomial on the argument $d$,
but in a situation in which the coefficients
$a_0,\ldots,a_{s-1}$ of the polynomial are
available only in the form of the integer $x$.

First one can observe that the problem is
well-defined:
Because $c\ge 2$ and $0\le a_j<c$ for
$j=0,\ldots,s-1$,
the mapping from $(a_0,\ldots,a_{s-1})$
to $\sum_{j=0}^{s-1}a_j c^j$ is injective,
except that the mapping is insensitive
to trailing zeros in its argument, so
$y$ is uniquely determined by~$x$.

Let $f=\Tceil{\log_2\max\{c,d\}}$ and take
$q=2+{{s f}/w}$.
Thus $q$ is essentially the number of $w$-bit words
occupied by $x$ and $y$.
Compute $t$ as the smallest positive integer with $c^{2^t}>x$.
Using repeated squaring, $t$ can be obtained
in $O(\sum_{k=0}^{t-1}\Tceil{2^{-k}q\log(2+2^{-k}q)})=O(t+q\log q)$ time
according to Lemma~\ref{lem:multiple}.
By adding or dropping trailing zeros as appropriate,
we can assume that $s=2^t$.
We will convert from $c$ to $d$ via $2^f$ in the sense
of first computing $z=\sum_{j=0}^{s-1}a_j 2^{f j}$
from $x$ and subsequently obtaining $y$ from $z$.
The significance of $f$ is that in the binary representation of
$z$ the coefficients $a_0,\ldots,a_{s-1}$ are readily available as the
contents of $s$ \emph{fields} of $f$ bits each.
This makes it easy to compute $y$ from $z$ via a
word-parallel version of a straightforward
divide-and-conquer procedure.
For $k=0,\ldots,t-1$, the procedure partitions $s/{2^k}$ digits
to base $b=d^{2^k}$ into pairs of consecutive digits and
replaces each pair
$(a',a'')$ by a single digit to base $b^2=d^{2^{k+1}}$
with the same value, namely $a'+b a''$.
Using word parallelism, this can be formulated as follows:

\begin{tabbing}
\quad\=\quad\=\quad\=\quad\=\kill
\>$b:=d$; $(*$ the current base $*)$\\
\>$r:=f$; $(*$ the current region size $*)$\\
\>$\widetilde y:=z$; $(*$ interpreted according to $b$ and $r$,
 $\widetilde y$ always has the value $y$ $*)$\\
\>\textbf{for} $k:=0$ \textbf{to} $t-1$ \textbf{do}\\
\>\>$(*$ $b=d^{2^k}$ and $r=2^k f$ $*)$\\
\>\>$u:=(2^r-1)\cdot 1_{{{s f}/{(2 r)}},2 r}$; $(*$ bit mask for keeping every other region $*)$\\
\>\>$\widetilde y:=(\widetilde y\,\mathbin{\textsc{and}}\,u)
 +((\widetilde y\gg r)\,\mathbin{\textsc{and}}\,u)\cdot b$; $(*$ two digits are combined into one $*)$\\
\>\>$b:=b^2$;\\
\>\>$r:=2 r$;
\end{tabbing}

\noindent
The final value of $\widetilde y$ is $y$.
To understand the code, note that $u$ is computed as a bit mask
with the property that forming the conjunction with $u$ picks out
every other \emph{region} of $r=2^k f$ bits, starting with
the least significant one.
Regions hold the digits to base $b=d^{2^k}$ alluded to above.
They start as single fields of $f$ bits
and double in size in every iteration.
At the end, there is only a single region of $s f$ bits that
contains the integer $y$.
The computation of $y$ takes
$O(q\sum_{k=0}^{t-1}\log(2+2^{-k}q))=O(q(t+(\log q)^2))$ time.

Suppose that $p\in\TbbbN$ and that $p$
instances of the problem just solved
have the same value of $d$ but different values
$z_1,\ldots,z_p$ of~$z$ and that $z_1,\ldots,z_p$
are presented as a single sequence of $p s f$ bits
in consecutive regions of $s f$ bits each,
where $s$ and $f$ are now given and $f$ is
at least $\Tceil{\log_2\max\{c,d\}}$.
If we again take $q=2+{{s f}/w}$,
the procedure is easily modified to solve all the instances
simultaneously
in $O(\Tceil{{{p s f}/w}}(t+(\log q)^2))$ time.
Only two aspects need attention.
First, the mask $u$ must be extended to ``cover'' all instances.
Second, the assumption that $s$ is a power of~2 may cause
an instance to ``encroach on'' its left neighbor.
To counter this, one can simply solve the even-numbered
instances in a first round and the odd-numbered instances
in a subsequent round.

The procedure for obtaining $z$ from $x$ is essentially
the reverse of the procedure for obtaining $y$ from~$z$:
For $k=t-1,\ldots,0$, each of $s/{2^{k+1}}$ digits $a$
to base $b^2=c^{2^{k+1}}$ is split into the two
digits $a'=a\bmod b$ and $a''=\Tfloor{a/b}$
to base $b=c^{2^k}$ and replaced by
the pair $(a',a'')$.
Of course, it is easy to obtain $a'$ from $a''$, even
in a word-parallel setting, as $a'=a-b a''$, 
but the formula for $a''$ involves division, which is not
in general readily amenable to word parallelism.
If all divisors are the same integer $b\ge 1$, however, as
is the case here, division by $b$ can be replaced by
multiplication by its approximate inverse.
The details are worked out in the following lemma.

\begin{lemma}
\label{lem:division}
Given integers $b,r\ge 1$ and an integer $x$ of the form
$x=\sum_{j=0}^{p-1} a_j 2^{r j}$, where $p\in\TbbbN$ and
$a_0,\ldots,a_{p-1}$ are integers with
$0\le a_j<2^r$ for $j=0,\ldots,p-1$,
the quantity
$\sum_{j=0}^{p-1}\Tfloor{{{a_j}/b}} 2^{r j}$
can be computed in
$O(\Tceil{{{p r}/w}}\log(2+{r/w}))$ time
with $O(p r+w)$ bits of working memory.
\end{lemma}

\begin{proof}
We first argue that for all integers $a\ge 0$ and
$t>a^2$,
$\Tfloor{{a/b}}=\Tfloor{{{a\cdot\Tceil{{t/b}}}/t}}$.
If $b>a$, the left-hand size is zero, and the right-hand
size is also zero, since
\[
\frac{a}{t}\cdot\left\lceil\frac{t}{b}\right\rceil
\le\frac{a}{t}\cdot\left\lceil\frac{t}{a+1}\right\rceil
\le\frac{a}{t}\cdot\frac{t+a}{a+1}
=\frac{a t+a^2}{a t+t}<1.
\]
If $b\le a$ and hence ${a/t}\le{1/b}$,
\[
\frac{a}{b}\le\frac{a}{t}\left\lceil\frac{t}{b}\right\rceil
<\frac{a}{t}\left(\frac{t}{b}+1\right)
=\frac{a}{b}+\frac{a}{t}\le\frac{a+1}{b},
\]
and there are no integers strictly between $a/b$ and ${{(a+1)}/b}$.

If there would be no interference between regions of $r$ bits,
the regionwise division by $b$ could be carried out according
to the formula
$\Tfloor{{a/b}}=\Tfloor{{{a\cdot\Tceil{{t/b}}}/t}}$,
used with $t=2^{2 r}$, simply by multiplying $x$
by $\Tceil{{t/b}}$, shifting the result right by
$2 r$ bits, and removing unwanted bits with a mask.
There is interference between regions, but because
$a_j\cdot\Tceil{{t/b}}<2^{3 r}$ for $j=0,\ldots,p-1$,
it suffices to carry out the computation in three
rounds, with each round involving every third region.
The time bound again follows from Lemma~\ref{lem:multiple}.
\end{proof}

In light of Lemma~\ref{lem:division}, it is easy to see
that $z$ can be computed from $x$ in
$O(q(t+(\log q)^2))$ time and,
again, that several instances with a common $c$
and different values of $z$ presented together
in $O(p s f)$ bits
can be solved simultaneously in
$O(\Tceil{{{p s f}/w}}(t+(\log q)^2))$ time.
The same therefore holds for the complete
computation of $y$ from~$x$.
Formally, we can express our findings as follows.

\begin{lemma}
\label{lem:base}
Given positive integers $c$, $d$, $f$ and $s$ with
$c,d\ge 2$ and $f\ge\Tceil{\log_2\max\{c,d\}}$
and an integer of the form
$\sum_{i=0}^{p-1}\left(\sum_{j=0}^{s-1}a_{i,j}c^j\right)\cdot 2^{i s f}$,
where $p\in\TbbbN$ and $0\le a_{i,j}<\min\{c,d\}$
for $i=0,\ldots,p-1$ and $j=0,\ldots,s-1$,
the integer
$\sum_{i=0}^{p-1}\left(\sum_{j=0}^{s-1}a_{i,j}d^j\right)\cdot 2^{i s f}$
can be computed in
$O(\Tceil{{p s f}/w}(\log s+(\log(2+{{s f}/w}))^2))$ time
with $O(s f+w)$ bits of working memory.
\end{lemma}

\subsection{A Small Power-of-2-Colored Choice Dictionary}
\label{subsec:small}

This subsection describes a
power-of-$2$-colored choice dictionary $D$
that is very slow for all but
the smallest universe sizes.
It is the core building block of the more
generally useful power-of-2-colored choice
dictionary presented in the next subsection.
We always assume that $c=2^{O(w)}$, so that
colors can be manipulated in constant time.
Instead of \Tvn{choice},
$D$ supports the
operation \Tvn{successor}, defined as follows,
where
$(S_0,\ldots,S_{c-1})$ is $D$'s client vector.

\begin{description}
\item[\normalfont$\Tvn{successor}(j,\ell)$]
($j\in\{0,\ldots,c-1\}$ and $\ell$ is an integer):
With $I=\{i\in S_j\mid i>\ell\}$,
returns $\min I$ if $I\not=\emptyset$, and 0 otherwise.
\end{description}

Recall that we call $D$'s client vector
\emph{deficient} if $S_j=\emptyset$ for at least
one $j\in\{0,\ldots,c-1\}$, i.e., if some color
is entirely absent.
If the client vector is not deficient,
it is \emph{full}.
The main feature of
$D$ is that it needs
less space when its client vector is deficient.
Although Lemma~\ref{lem:atomic} below
mentions \Tvn{successor} instead of
\Tvn{choice}, we still speak of a choice dictionary
because \Tvn{choice} reduces to \Tvn{successor}
(instead of $\Tvn{choice}(j)$, execute
$\Tvn{successor}(j,0)$).

\begin{lemma}
\label{lem:atomic}
There is an externally sized power-of-2-colored
choice dictionary $D$ that, for arbitrary given
$m,f\in\TbbbN$,
can be initialized for
universe size $m$ and $c=2^f$ colors in
$O(\Tceil{{m f}/w})$ time and
subsequently needs to store in an external
\emph{fullness} bit
whether its client vector is full, executes
\Tvn{color} in
$O({{m f}/w}+\Tceil{{c f}/w}\Lambda)$ time
and \Tvn{setcolor} and \Tvn{successor} in
$O(\Tceil{{m f}/w}\Lambda)$ time,
where $\Lambda=f+(\log(2+{{c f}/w}))^2$,
and occupies at most $m f$ bits
and at most $m f-{m/c}+2 c f$ bits
when its client vector is deficient.

Alternatively, if initialized with an additional
parameter $t\in\TbbbN$ and given access to suitable tables
of at most $c^{c/t}$ bits that can be computed
in $O(c^{c/t})$ time and depend only on $c$ and~$t$,
$D$ can execute \Tvn{color}
in $O({{mf}/w}+\Tceil{{{c f}/w}}t)$ time.

The transient space needed by $D$ is $O(m f+w)$ bits.
\end{lemma}

\begin{proof}
We view $D$'s task as that of maintaining a
sequence
of $m$ color values or \emph{digits} drawn from
the alphabet
$\Sigma=\{0,\ldots,c-1\}$, where $c=2^f$.
When its client vector is full,
$D$ employs a \emph{standard representation}
that stores
the $m$ color values as the concatenation of their
binary representations, each of which is given in
an $f$-bit field.
Assume first that $D$ is in the standard representation.
The operation \Tvn{successor} must locate the
first occurrence, if any, of a particular color $j$
after a certain position.
This can be carried out in $O(\Tceil{{{m f}/w}})$
time with the algorithm of Lemma~\ref{lem:log}(b)
after forming the \texttt{xor} with
$j\cdot 1_{m,f}$.
It is trivial to execute \Tvn{color}
in constant time by inspecting
the value of a single field.
Similarly, \Tvn{setcolor} updates the value
of a single field.
If this makes the client vector deficient, 
however, $D$ is converted to a
\emph{compact representation} described in
the following.

We will assume that $m$ is a multiple of $c^2 f$,
noting that up to $c^2 f-1$ 
``surplus'' digits, always kept in the standard
representation, can be handled
within the time bounds of the lemma,
as argued in the previous paragraph.
The compact representation partitions the $m$
digits into \emph{groups} of $c$ consecutive
digits each.
The compact representation also stores
the bit vector $Z=(z_0,\ldots,z_{c-1})$,
where $z_j=1$, for $j=0,\ldots,c-1$, exactly if
$S_j\not=\emptyset$.
Since $Z$ is easy to initialize when $D$ is
converted to the compact representation and
subsequently can change only in a call of \Tvn{setcolor},
and then only in at most
two bits whose values can be tested with 
\Tvn{successor}, maintaining $Z$
is not a bottleneck.
As already mentioned,
the compact representation is used only when
$J_0=\{j\in\{0,\ldots,c-1\}\mid z_j=0\}$ is nonempty.
When this is the case, $j_0=\min J_0$ can be
computed in $O(\Tceil{{c/w}})$ time with the algorithm
of Lemma~\ref{lem:log}(b).
Let $\Tvn{skip}_{j_0}$ be the increasing bijection
from $\Sigma\setminus\{j_0\}$ to $\Sigma'=\{0,\ldots,c-2\}$.

The conversion from the standard representation
to the compact representation
is done in three successive steps:
\emph{excision}, \emph{base change}, and
\emph{compaction}.
The excision excludes the unused color $j_0$
from the alphabet
by applying $\Tvn{skip}_{j_0}$ independently to
each of the $m$ digits.
This can be done in $O(\Tceil{{{m f}/w}})$ time
as described in \cite{HagK16}:
First the algorithm of
Lemma~\ref{lem:log}(c)
is used to compute an
integer $y$, each of whose fields---with
$y$ viewed as composed of $m$ fields of $f$ bits each---stores~1
if the corresponding digit is $\le j_0$,
and~0 otherwise.
The application of $\Tvn{skip}_{j_0}$ to all
digits is finished by subtracting $1_{m,f}-y$
from the
$(m f)$-bit standard representation,
viewed as a single integer.
Now we have a sequence of $m$ transformed digits
drawn from the smaller alphabet $\Sigma'$,
but still stored in $f$-bit fields.

Each group, composed of the (transformed) digits
$a_0,\ldots,a_{c-1}$, say, can be viewed as
representing the integer $\sum_{j=0}^{c-1}a_j c^j$.
The base change uses
the algorithm of Lemma~\ref{lem:base} with $s=c$
to replace $\sum_{j=0}^{c-1}a_j c^j$
by $\sum_{j=0}^{c-1}a_j (c-1)^j$ independently
within each group.
This takes
$O(\Tceil{{{m f}/w}}\Lambda)$ time, where
$\Lambda=f+(\log(2+{{c f}/w}))^2$, and,
informally, encodes each group more economically.
Indeed, since
$c\log_2(c-1)=c\log_2\! c+c\log_2(1-{1/c})
\le c f+c\ln(1-{1/c})\le c f-1$,
within each group of $c f$ bits that hold
a group the most significant bit is~\texttt{0}.

At this point the entire representation,
viewed as an integer $u$, consists
of $h={m/c}$ repetitions of a pattern
consisting of
$g-1=c f-1$ bits considered to be in use followed
by a single bit that is unused,
and $h$ is a multiple of~$g$.
In order to satisfy the space bound of the lemma,
the compaction reorders the
$h(g-1)$
used bits in $u$ and stores them
tightly in the $h(g-1)=m f-{m/c}$
least significant bit positions.
This can be done in a way
illustrated in Fig.~\ref{fig:6}:
The part $v$ of $u$ consisting of its
least significant
$r=h-1$ bits,
whose
$s=h-{h/g}$ used bits 
are labeled $1,\ldots,12$ in the figure,
is broken off and replicated
$g-1$ times
through a multiplication with
$1_{g-1,r}$
to yield an integer~$x$.
The remaining larger part of $u$
is shifted right by $r$ bits to yield an integer~$y$.
Now the positions in~$x$ of the
$s$ unused bits in $y$
in least significant positions
(i.e., ignore the leading unused bit of~$y$)
together hold copies of all $s$ used bits
in $v$,
so that applying a suitable mask to $x$
and adding the result to~$y$
finishes the computation.
The steps just
described consist in evaluating the expression
$
(u\gg r)+(((u\,\mathbin{\textsc{and}}\,(2^r-1))\cdot 1_{g-1,r})
 \,\mathbin{\textsc{and}}\,1_{s,g})
$,
which can be done in $O(\Tceil{{{m f}/w}})$ time.

\begin{figure}
\begin{center}
\epsffile{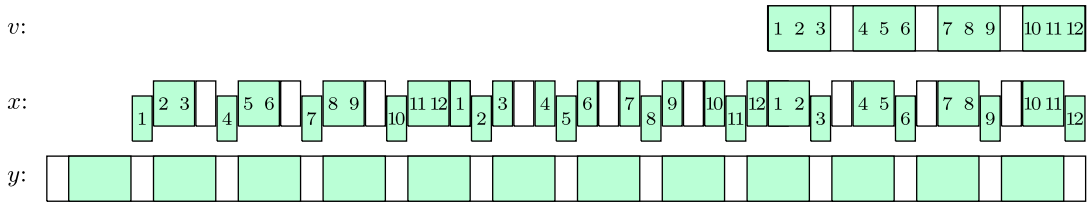}
\caption{The compaction of the used bits in an example with
$g=4$ and $h=16$.}
\label{fig:6}
\end{center}
\end{figure}

In order to convert the compact representation back
to the standard representation,
we reverse the steps described above.
The integer $v$ can be restored
by extracting
its $s$ used bits with a mask, multiplying them
by $1_{g-1,r}$, shifting the result right by
$(g-2)r$ bits and forming the conjunction
with $2^r-1$.
After concatenating $v$ with $y$ to obtain~$u$,
we use the algorithm of Lemma~\ref{lem:base}
to replace $\sum_{j=0}^{c-1}a_j (c-1)^j$
by $\sum_{j=0}^{c-1}a_j c^j$ within each group.
Finally $\Tvn{skip}_{j_0}^{-1}$ can be applied
independently to each digit
much as $\Tvn{skip}_{j_0}$ was.
The entire conversion from the standard to the
compact representation or back takes
$O(\Tceil{{m f}/w}\Lambda)$ time.

When $D$ is in the compact representation,
we can execute
\Tvn{color} by undoing the compaction
($O(\Tceil{{{m f}/w}})$ time), undoing the base change
and the excision for the single relevant group
($O(\Tceil{{c f}/w}\Lambda)$ time),
and finally reading out the $f$ bits of interest.
To execute \Tvn{setcolor} and
\Tvn{successor}, we carry out the complete
conversion from the compact to the standard representation,
apply the corresponding
algorithm for the standard representation and,
in the case of \Tvn{setcolor} and
unless an inspection of (a saved copy of)
$Z$ shows that $D$'s
client set has become full, convert
$D$ back to the compact representation.
This takes $O(\Tceil{{m f}/w}\Lambda)$ time.

Because $m$ may not actually be a multiple
of~$c^2 f$, the number of bits saved by the
compact representation relative to the standard
representation is not necessarily $m/c$, but still
at least ${{(m-c^2 f)}/c}$.
Moreover, $c$ bits are needed for the bit vector~$Z$.
Hence the number of bits occupied by the compact
representation is at most
$m f-{m/c}+2 c f$, as indicated in the lemma.

For given $t\in\TbbbN$, suitable tables of at most
$c^{{c/t}}$ bits that can be computed in
$O(c^{{c/t}})$ time and depend only on $c$ and~$t$
allow us to carry out a base change in a
group in $O(t)$ time via table lookup
(say,
$\Tceil{c/{(2 t)}}$ digits at a time).
The alternative time bound
for \Tvn{color} follows easily.
\end{proof}

\subsection{An Unrestricted Power-of-2-Colored Choice Dictionary}
\label{subsec:general2}

The top-level idea behind the colored choice dictionary
of this section is to keep the overall organization
of the uncolored choice dictionary of
Section~\ref{sec:uncolored}, but now letting weak
indices correspond to deficient client vectors
implemented with the data structure of
Lemma~\ref{lem:atomic} in order to gain the space
needed for pointers to mates.

\begin{theorem}
\label{thm:colored2}
There is an externally sized
power-of-$2$-colored choice dictionary $D$
that, for arbitrary given $n,f\in\TbbbN$,
can be initialized for universe size $n$
and $c=2^f$ colors in constant time
and subsequently occupies $n f+1$ bits and
executes \Tvn{color} in
$O({{c(c f+\log n)f}/w}+\Tceil{{c f}/w}\Lambda)$ time
and \Tvn{setcolor} and \Tvn{choice} in
$O(c\Tceil{{(c f+\log n)f}/w}\Lambda)$ time,
where $\Lambda=f+(\log(2+{{c f}/w}))^2$.
In particular, if $c f=O(w)$,
\Tvn{color} runs in
$O({{(c f\log n)}/w}+f)=O(c f)$ time and
\Tvn{setcolor} and \Tvn{choice} run in $O(c f^2)$ time.
For constant $c$ the choice dictionary is atomic.

Alternatively, if initialized with an additional
parameter $t\in\TbbbN$ and given access to suitable tables
of at most $c^{c/t}$ bits that can be computed
in $O(c^{c/t})$ time and depend only on $c$ and~$t$,
$D$ can execute \Tvn{color} in
$O({{c(c f+\log n)f}/w}+\Tceil{{c f}/w}t)$ time.

The transient space needed by $D$ is $O(c(c f+\log n)f+w)$ bits.
\end{theorem}

\begin{proof}
For the time being ignore the claim
about constant-time initialization.
Concerning many aspects, described in this paragraph, the
colored choice dictionary $D$
of Theorem~\ref{thm:colored2}
is similar to the uncolored choice dictionary
of Section~\ref{sec:uncolored}.
Consider the situation following an initialization
of $D$ for universe size $n$ and $c$ colors and
assume first that $n$ is a multiple
of an integer $N\in\TbbbN$ that will be chosen later.
Most of $D$'s information is kept in an array $A$ of
$N$ cells $A[1],\ldots,A[N]$
that we now call \Tvn{containers}.
Correspondingly, we view $D$'s task
as that of maintaining a sequence $(a_1,\ldots,a_N)$,
where $a_k$ is a sequence of $m=n/N$ color values
drawn from $\{0,\ldots,c-1\}$,
for all $k\in V=\{1,\ldots,N\}$.
Again $D$ stores an integer
\emph{barrier} $\mu$ with $0\le \mu\le N$,
and an integer $k\in V$ is said
to be to the left of the barrier if $k\le \mu$
and to its right otherwise.
An integer $k$ to the left of the barrier is
\emph{matched} to an integer $\ell$ to the right
of the barrier exactly if a designated field
in $A[k]$,
called $A[k].\Tvn{mate}$,
contains $\ell$ and
$A[\ell].\Tvn{mate}$ contains~$k$, and then $k$ and
$\ell$ are \emph{mates}.
As in Section~\ref{sec:uncolored}, let
$\Tvn{mate}(k)$ be the mate of $k$
if $k$ is matched and $k$ itself if not,
for all $k\in V$.
Again, $k\in V$ is \emph{strong} if $k$ is matched
and $k\le \mu$ or $k$ is unmatched and $k>\mu$, and
$k$ is \emph{weak} if it is not strong.
For convenience, we will apply the terms of
being to the left or right of the barrier,
matched, mates, strong and weak also to containers,
saying that $A[k]$ is to the left of the barrier
exactly if $k$ is, for all $k\in V$, etc.
If a container $A[k]$ is strong and unmatched,
it simply stores $a_k$ as a sequence of $m f$ bits.
If $A[k]$ is strong and matched to $A[\ell]$,
$A[k]$ stores the biggest part,
$\underline{a_k}$,
of $a_k$
and the rest of $a_k$,
$\overline{a_k}$, is stored in a field
$A[\ell].\Tvn{top}$ of~$A[\ell]$.
The part of $A[k]$ not taken up by $\underline{a_k}$
holds $A[k].\Tvn{mate}$ as well as
an \emph{auxiliary field}
$A[k].\Tvn{aux}$ of $O(c+\log n)$ bits.
Similarly as in the construction of
Subsection~\ref{subsec:hiding}, we store the barrier
$\mu$ in $A[1].\Tvn{aux}$ (except if $\mu=0$).
Again, the similarity to the data organization
of the uncolored choice dictionary is pronounced.

The most significant difference to the situation
in Section~\ref{sec:uncolored} is that if
$k\in V$
is weak, we can no longer conclude that $a_k$ is
zero (or a sequence of zeros) and hence that no
information must be stored about $a_k$ beyond the
fact that $k$ is weak.
Instead the convention here is that $A[k]$ is
weak exactly if (the client vector corresponding
to) $a_k$ is deficient, i.e., if some color does
not occur in~$a_k$.
Thus even if $A[k]$ is weak, it must store information
``of its own'', but the deficiency of~$a_k$
makes it possible to do this in less space.

We realize each container $A[k]$ as
an instance of the data structure of
Lemma~\ref{lem:atomic}, initialized for universe
size $m$ and $c$ colors.
Thus a container $A[k]$ needs at least
${m/c}-2 c f$ fewer bits when its client
vector is deficient than when it is full.
Informally, we can express this by saying that
$A[k]$ can carry a \emph{payload} of
at least ${m/c}-2 c f$ bits when its
client vector is deficient---so many unrelated
bits can be stored within the space
reserved for~$A[k]$.
We need containers to have a payload of
at least $K(c f+\log n)$ bits for some constant
$K\in\TbbbN$ and achieve this by choosing
$m=\Theta(c(c f+\log n))$ appropriately.

If $k\in V$ is strong, $A[k]$ is in the standard
representation.
It has no payload, but
$O(c+\log n)$ of its bits
form the fields $A[k].\Tvn{mate}$ and
$A[k].\Tvn{aux}$.
As mentioned above,
if $k$ is strong and matched, the rest of $A[k]$ stores
$\underline{a_k}$.
If $k$ is weak, $A[k].\Tvn{mate}$, $A[k].\Tvn{aux}$ and
$A[k].\Tvn{top}$
constitute the payload of $A[k]$.
The logical realization of
the fields
\Tvn{mate} and \Tvn{aux}
can be seen to be different in weak and strong
containers, but we ensure that they are
located in the same
bits in the two cases so that,
in particular, it can be determined
in constant time whether a container is
weak or strong.
This realizes in a procedural way the external
fullness bit required by Lemma~\ref{lem:atomic}.

Each container also contributes
$c$ \emph{special bits}, one to each of
$c$ dynamic uncolored choice dictionaries
$D_0,\ldots,D_{c-1}$ whose universe sizes are
kept equal to $\mu$ at all times.
One may think of the special bits of a
container as located
in the container, but in fact $D_0,\ldots,D_{c-1}$
are stored in contiguous memory locations,
and whenever a container wants to inspect or
change one of its $c$ special bits, it must
call the appropriate operation in one of
$D_0,\ldots,D_{c-1}$.
We shall say that $D_0,\ldots,D_{c-1}$ are
\emph{distributed} over $A[1],\ldots,A[N]$.
Each of $D_0,\ldots,D_{c-1}$ needs one
additional bit, which is stored in $A[1].\Tvn{aux}$
(except if $\mu=0$, in which case
the states of $D_0,\ldots,D_{c-1}$ can be arbitrary).
The positions of the $c$ special bits in a
container are chosen
within the payload of the compact
representation, but outside of the parts
of the payload used for other purposes, and
such that in the standard representation of the
container no single
color value is stored in bits that include two
or more special bits; this is possible with a
payload of $c f$ bits reserved for this purpose.
The operations are extended to
maintain as an invariant for $j=0,\ldots,c-1$
that an integer $k\in\{1,\ldots,\mu\}$
belongs to the client set of $D_j$
exactly if $j$ occurs as a color value in~$a_k$.
Correspondingly, the client sets of $D_1,\ldots,D_{c-1}$
are initialized to be empty, whereas the initial client set
of $D_0$ is the entire set $V$.

The information of its own that $A[k]$ stores
when $k$ is weak
(i.e., what ``pays for'' the entire payload)
is (the deficient) $a_k$.
An unmatched weak container $A[k]$ has the same structure
as a matched weak container, except that 
$A[k].\Tvn{mate}$ and
$A[k].\Tvn{top}$ are arbitrary---$A[k].\Tvn{mate}$
may not give rise to a spurious matching edge,
though.

To execute \Tvn{color}, we
must determine a single color value in $a_k$
for some $k\in V$.
Comparing $k$ to $\mu$ and
inspecting $A[k].\Tvn{mate}$ and
possibly $A[\ell].\Tvn{mate}$ for some $\ell\in V$, we can
discover in constant time whether $A[k]$
is matched and, if so, its mate.
This allows us to identify the container that
contains the relevant color value, and we finish
by returning the value obtained
by calling \Tvn{color} for that container
with an appropriate argument,
which may require us to retrieve a special bit.
Since $m=\Theta(c(c f+\log n))$, the operation takes
$O({{c(c f+\log n)f}/w}+\Tceil{{c f}/w}\Lambda)$ time.

To execute $\Tvn{choice}(j)$, we distinguish between two cases.
If $\mu=N$,
we compute $k=D_j.\Tvn{choice}$ and return~0
if $k=0$ and
$(k-1)m+A[k].\Tvn{successor}(j,0)$ otherwise.
If $\mu<N$, the color $j$ occurs in $a_k$,
where $k=\Tvn{mate}(N)$, and a suitable return
value can be obtained by executing
$\Tvn{successor}(j,0)$ in $A[k]$ and possibly
in the mate of $A[k]$.
We may have to retrieve up to $2 c$ special bits,
so the total time comes to
$O(c\Tceil{{{(c f+\log n)f}/w}}\Lambda)$.

To execute $\Tvn{setcolor}(j,\ell)$, 
first read out the old color
$j_0=\Tvn{color}(\ell)$ of~$\ell$.
Assume that $j\not=j_0$.
Determine the $k\in V$ such that the color
of~$\ell$ is a component of $a_k$
and take $k'=\Tvn{mate}(k)$.
In the following, by ``eliminating a possible
spurious matching edge at $i$'', where $i\in V$,
we mean the following:
If $i$ has a mate $i'$, then change this
fact by setting $A[i'].\Tvn{mate}$ to $i'$.

If $k$ is weak, save the payload of $A[k]$
before changing the color of $\ell$ from
$j_0$ to $j$ through an appropriate call
of \Tvn{setcolor} in $A[k]$.
If the client set of $A[k]$
continues to be deficient,
i.e., if $A[k]$ does not attempt to change its
fullness bit,
nothing more needs to be done, except that if $k\le\mu$,
$D_{j_0}$ and $D_j$
should be updated
appropriately with respect to~$k$
(again, the necessary tests can
be carried out with \Tvn{successor}).
If the client set of $A[k]$ becomes full,
proceed to carry out what
corresponds to an insertion
in Section~\ref{sec:uncolored}.
Take $\tilde{\mu}=\mu$, compute
$\mu'=\Tvn{mate}(\mu)$ and decrease $\mu$ by~1.
If $\tilde{\mu}\not=\mu'$, i.e., if $\tilde{\mu}$
is strong, then restore
$a_{\tilde{\mu}}$ by storing
$\overline{a_{\tilde{\mu}}}=A[\mu'].\Tvn{top}$ in
the appropriate bits of~$A[\tilde{\mu}]$
(if $\mu'=k$, this involves the saved
payload of $A[k]$) and
eliminate a possible spurious matching edge at~$\tilde{\mu}$.
If $k\not=\mu'$, set $A[\mu'].\Tvn{top}:=A[k].\Tvn{top}$
and match $k'$ and $\mu'$ by executing
$A[k'].\Tvn{mate}:=\mu'$ and
$A[\mu'].\Tvn{mate}:=k'$.
Finally
if $k>\mu$, eliminate a
possible spurious matching edge at~$k$.

If $k$ is strong, first change the color of $\ell$ from
$j_0$ to $j$ through an appropriate call of \Tvn{setcolor}
in a container.
Then determine with one or two calls of
$\Tvn{successor}(j_0,0)$
whether the color $j_0$ still occurs in~$a_k$.
If it does, nothing more needs to be done.
Otherwise
proceed to carry out what
corresponds to a deletion
in Section~\ref{sec:uncolored}.
If $k\le\mu$, overwrite
the appropriate part of $A[k]$
with $A[k'].\Tvn{top}$.
Let $\mu'=\Tvn{mate}(\mu+1)$ and increase $\mu$ by~1.
If $k\le\mu$, remove $k$ from $D_{j_0}$.
If $\mu'\not=k$,
observe that $\overline{a_{\mu'}}$ is stored in
$A[\mu].\Tvn{top}$ if $\mu\not=\mu'$
and as part of the value of $A[\mu']$ otherwise
and save $\overline{a_{\mu'}}$ in a variable~$v$.
Determine for each $j'\in\{0,\ldots,c-1\}$
with a call of $A[\mu].\Tvn{successor}(j,0)$
whether the color $j'$ occurs in $a_\mu$
and ensure that $\mu$ belongs to the client set
of $D_{j'}$ if and only if this is the case.
Then execute $A[k'].\Tvn{mate}:=\mu'$ and
$A[\mu'].\Tvn{mate}:=k'$, which matches $k'$ and $\mu'$
except if $k'$ and $\mu'$ are on the same
side of the barrier.
Finally, if $\mu'\not=k$, restore $a_{\mu'}$
by executing $A[k'].\Tvn{top}:=v$.

Since the total number of operations executed on containers
and the number of special bits that need to retrieved
and written back are both $O(c)$,
\Tvn{setcolor} can be seen to operate within the
time bound of
$O(c\Tceil{{c(c f+\log n)f}/w}\Lambda)$
indicated in the theorem.
In order to satisfy the assumption that $n$ is a
multiple of $N$, we maintain $O(c(c f+\log n)f)$
surplus digits in a single instance of the
data structure of Lemma~\ref{lem:atomic} that is
always kept in the standard representation.
The time bounds
of the theorem can still be guaranteed.

A final issue to be addressed is the constant-time
initialization of~$D$.
Viewing $D$ as composed of $w$-bit words
(plus possibly one incomplete word that can be
initialized in constant time), we can almost
provide the initialization using the
initializable arrays of Katoh and Goto~\cite{KatG17},
but need to modify them in two ways.
First, Katoh and Goto consider the initialization
of all array entries to the same value $v$, but here
we need an initialization of $D$ to a bit pattern in
which all colors are 0, the client
sets of $D_1,\ldots,D_{c-1}$ are empty, and the
client set of $D_0$ is $\{1,\ldots,N\}$.
We can handle this issue using a simple mechanism,
described by Hagerup and Kammer~\cite{HagK17},
that consists in letting a fixed value $v$
represent a
``word-sized slice'' of the desired initial bit pattern,
while conversely using the slice to represent~$v$.
The slice depends on the position in the array,
but is easy to compute from that position.
Second, the data structure of Katoh and Goto needs
a bit $\Tvn{flag}$ in addition to the bits of the array that
it maintains, $\Tvn{flag}=1$ signifying that all
positions in the array have been written to,
much as a special bit is used in
Subsection~\ref{subsec:hiding} to signify
that $\mu=0$.
Here we have already used all of
the $n f+1$ bits allowed by Theorem~\ref{thm:colored2}
and have no bit to spare.
As long as $\mu>0$, however, \Tvn{flag}
can be stored in $A[1].\Tvn{aux}$.
Moreover, without violating the time
bound of \Tvn{setcolor}
we can easily ensure that all words of
all containers to the right of the barrier have
been written to, which implies that \Tvn{flag}
is superfluous (its value is known to be~1)
whenever $\mu=0$.
Thus maintaining \Tvn{flag} does not cost
any extra space.

The alternative bound for \Tvn{color} is obtained simply by
appealing to the corresponding part of
Lemma~\ref{lem:atomic}.
All containers can share the same tables.
\end{proof}

The alternative time bounds of Theorem~\ref{thm:colored2}
depend on an external table.
As expressed in the following theorem,
we can also incorporate the table into
the data structure itself.

\begin{theorem}
\label{thm:colored2t}
There is an externally sized
power-of-$2$-colored choice dictionary
that, for\break
arbitrary given $n,f,t\in\TbbbN$,
can be initialized for universe size $n$
and $c=2^f$ colors in\break
constant time
and subsequently occupies
at most $n f+c^{c/t}+1$ bits and
executes \Tvn{color} in
$O({{c(c f+\log n)f}/w}+\Tceil{{c f}/w}t)$ time
and \Tvn{setcolor} and \Tvn{choice} in
$O(c\Tceil{{(c f+\log n)f}/w}\Lambda)$ time,
where $\Lambda=f+(\log(2+{{c f}/w}))^2$.
The transient space needed by $D$ is $O(w+c f\log n)$ bits.
\end{theorem}

\goodbreak

\begin{proof}
The theorem follows immediately from Theorem~\ref{thm:colored2},
except that we must show how to achieve a constant
initialization time despite the use of a table $Y$
of nontrivial size.

The main observation
is that before an entry in $Y$ is first needed,
with one exception,
it can be computed from an earlier entry
that it ressembles.
The reason is that $Y$ is used to map between
$\sum_{i=0}^{s-1} a_i c^i$ and $\sum_{i=0}^{s-1} a_i(c-1)^i$
for some $s\le c$, where $a_0,\ldots,a_{s-1}$ are
consecutive color values.
At the point where a new tuple $(a_0,\ldots,a_{s-1})$
of color values arises, it does so in a call of
\Tvn{setcolor}, and it is derived from a tuple
$(a'_0,\ldots,a'_{s-1})$ that differs from
$(a_0,\ldots,a_{s-1})$ in only one component.
The two entries for $(a_0,\ldots,a_{s-1})$
(one for each direction of the mapping)
can be computed from those of $(a'_0,\ldots,a'_{s-1})$
with a constant number of multiplications
and divisions by numbers of the form
$(c-1)^i$, with $i\in\{1,\ldots,s-1\}$.
Since two integers of at most $c f$ bits each
can be multiplied and divided in
$O(\Tceil{{{c f^2}/w}})$ time, it can be seen
that the computation can be accomplished within
the time bound for \Tvn{setcolor}
indicated in the theorem.
The first entries in $Y$, those corresponding to the tuple
$(0,0,\ldots,0)$, are trivial and can be filled in
in constant time during the initialization.
\end{proof}

\section{General Colored Choice Dictionaries}
\label{sec:colored}

We now turn from the case in which the number
$c$ of colors is a power of~2 to the case of
general $c\ge 2$, the immediate difficulty being
that a single color value cannot be stored in a number
of bits without an unacceptable waste of space.

\subsection{Compaction}

When $c$ is not a power of~2,
we need a compaction algorithm more general than
the one illustrated in Fig.~\ref{fig:6}.
It is characterized in the following lemma.

\begin{lemma}
\label{lem:compaction}
Given positive integers $n$, $m$ and $u$ with $u\le m$ and
an integer of the form
$\sum_{i=0}^{n-1}\sum_{j=0}^{u-1}2^{i m+j}b_{i u+j}$,
where $b_k\in\{0,1\}$ for $k=0,\ldots,n u-1$,
for a certain permutation $\sigma$ of
$\{0,\ldots,n u-1\}$ the integer
$\sum_{k=0}^{n u-1}2^k b_{\sigma(k)}$
can be computed in
$O(\Tceil{{{n m}/w}}\Lambda)$ time, where
$\Lambda=\log(\min\{u,m-u\}+2)$, using
$O(n m)$ bits of working memory.
Moreover, given $n$, $m$, $u$ and
$\sum_{k=0}^{n u-1}2^k b_{\sigma(k)}$,
the original integer
$\sum_{i=0}^{n-1}\sum_{j=0}^{u-1}2^{i m+j}b_{i u+j}$
can be reconstructed within
the same time and space bounds.
\end{lemma}

\begin{proof}
We provide only an informal proof sketch based
mostly on figures.
The positions of the bits $b_0,\ldots,b_{n u-1}$
can be visualized as what we will call a
\emph{group arithmetic progression}
with \emph{period} $m$, $n$ \emph{groups},
\emph{group size} $u$,
\emph{weight} $n u$ and
\emph{range} $n m$ (see Fig.~\ref{fig:8}).
Every group arithmetic progression considered in the following
has period~$m$ and range at most~$n m$
without this being stated explicitly.
If a group arithmetic progression has group size~$u$,
we call it a \emph{$u$-sequence}.
In Fig.~\ref{fig:8}, the vertical
bar is placed in \emph{position} $n u$,
i.e., with $n u$ positions to its right.
The task at hand
can therefore be viewed as that of mapping the balls
to the left of the bar bijectively
to the holes to the right of the bar.

\begin{figure}
\begin{center}
\epsffile{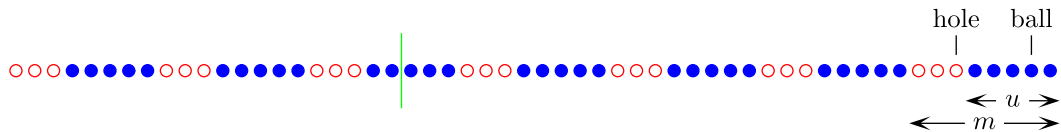}
\caption{The compaction problem: Mapping balls to holes.}
\label{fig:8}
\end{center}
\end{figure}

Imagine that the vertical bar splits into two copies.
One copy, the \emph{left bar}, moves left until it
hits the first position that is a multiple of~$m$.
The other copy, the \emph{right bar}, moves right
until it hits the first position that is a multiple
of $m$ and has the property that the number of holes
to its right is bounded by the number of balls
to the left of the left bar.
Ignoring for the time being the subproblem represented
by the balls and holes between the two bars, we
are faced with the problem of mapping a subset of the
$u_0$-sequence of balls to the left of the left bar
bijectively to the $v_0$-sequence of holes to the
right of the right bar, where $u_0=u$ and $v_0=m-u_0$.
More generally, we consider the problem of mapping
a subset of a \emph{left} $u$-sequence of balls bijectively
to a \emph{right} $v$-sequence of bins, where $u$ and $v$ are
positive integers with $u+v\le m$ and the weight of
the right sequence is bounded by that of
the left sequence
(see Fig.~\ref{fig:9}).

\begin{figure}
\begin{center}
\epsffile{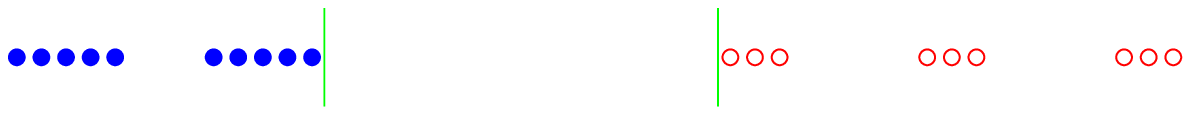}
\caption{Mapping one group arithmetic progression
 to another of no larger weight.}
\label{fig:9}
\end{center}
\end{figure}

If $u\le v$ (a group of balls fits in a group of holes),
we place some of the balls in some of the holes as
illustrated in Fig.~\ref{fig:10}, where groups of
balls are shown labeled consecutively in the order,
from right to left, in which they occur in the
left sequence.
Similarly as in the procedure of Fig.~\ref{fig:6},
this can be carried out in $O(\Tceil{{{n m}/w}})$ time,
mainly with a multiplication and a constant
number of bitwise Boolean operations and shifts.

\begin{figure}
\begin{center}
\epsffile{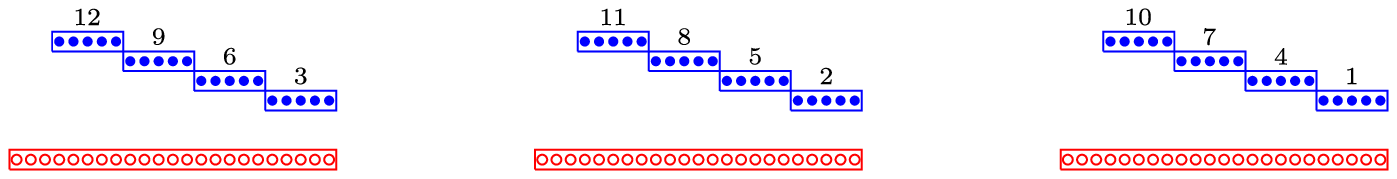}
\caption{A partial mapping of smaller groups of
 balls to larger groups of holes.}
\label{fig:10}
\end{center}
\end{figure}

If $u$ divides $v$, all holes are filled, and this
part of the computation is finished.
Otherwise what remains of the left sequence is still
a $u$-sequence (but with fewer groups), whereas the
remaining holes form a $(v\bmod u)$-sequence
(with the same number of groups).
We can therefore say that the computation of
Fig.~\ref{fig:10} reduces a $(u,v)$-instance
of the problem to a $(u,v\bmod u)$-instance.

If $u>v$ (a group of balls is larger than a group of holes),
we place some of the balls in some of the holes with
the alternative procedure shown in Fig.~\ref{fig:11},
which can again be carried out in $O(\Tceil{{{n m}/w}})$ time.
Here the condition $u+v\le m$ is essential, as it
prevents overlap between the different shifted copies
of the sequence of balls.
Sequences of $v$ balls are shown labeled by an integer
that indicates the group (of size $u$) of the left
sequence from which they originate and a letter
that indicates their position within that group.

\begin{figure}
\begin{center}
\epsffile{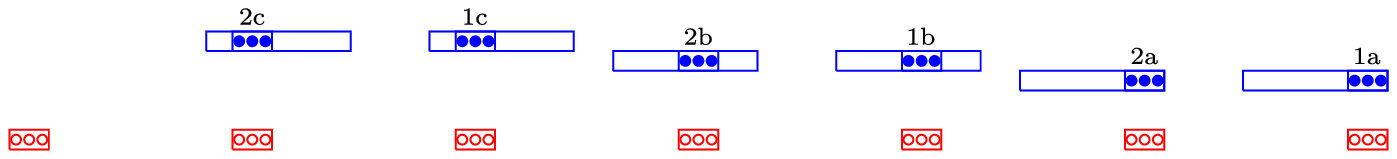}
\caption{A partial mapping of larger groups of
 balls to smaller groups of holes.}
\label{fig:11}
\end{center}
\end{figure}

If $v$ divides $u$, all holes are filled, and this
part of the computation is finished.
Otherwise what remains of the right sequence is
still a $v$-sequence (but with fewer groups),
whereas the balls that remain unplaced form a
$(u\bmod v)$-sequence
(with the same number of groups).
Thus the $(u,v)$-instance is reduced to a
$(u\bmod v,v)$-sequence.
It is well-known that the mapping that takes
$(u,v)$ to $(u,v\bmod u)$ if $u\le v$ and to
$(u\bmod v,v)$ if $u>v$, if started at $(u_0,v_0)$,
reaches a pair
with a zero component after $O(\log(\min\{u_0,v_0\}+2))$
repeated applications.
Indeed, if $u>v>0$, $u\bmod v\le{u/2}$.
This shows the running time claimed for
the compaction, except that we still have to
consider the ``middle'' instance ignored above.

Consider the balls of the middle instance that
are to the left of the original vertical bar
in Fig.~\ref{fig:8}, i.e., that are to be placed.
If their number is $p$, by construction, the
remaining holes form a $v_0$-sequence of
weight at most $p+v_0$.
Letting $p'$ be the largest integer bounded by~$p$
that is a multiple of $v_0$, we can use the method
of Fig.~\ref{fig:11} to place $p'$
consecutive of the $p$ balls in holes so that
the remaining holes form at most two groups
(that may not be of the same size).
In the same manner, the number of groups of
unplaced balls can be reduced below a constant with
the method of Fig.~\ref{fig:10} without increasing
the number of groups of remaining holes.
The remaining instance has $O(1)$ groups of balls
and holes and can therefore obviously
be solved in constant time.

It is not difficult to see that the steps
represented by Figs.\ \ref{fig:10} and~\ref{fig:11}
are reversible in the sense that the balls
placed in holes can be returned to their
original positions in $O({{n m}/w})$ time.
Therefore the whole computation is reversible
within the time bound of the lemma.
For the ``forward'' computation we must keep track
of a constant number of nonnegative integer
parameters that are functions of $n$, $m$ and $u$ and
bounded by $n m$.
In order to reverse the computation, we need
these parameters for every stage of the computation.
They can be obtained in $O(\Tceil{{m n}/w}\Lambda)$ time
by simulating the forward computation and
take up $O(\Lambda\log(n m+2))=O(n m)$ bits of memory.
\end{proof}

\subsection{A Small $c$-Color Choice Dictionary for General $c$}

\begin{figure}
\begin{center}
\epsffile{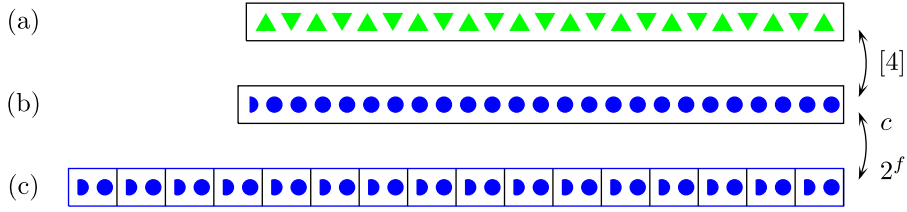}
\caption{A big digit in the representation
of~\cite{DodPT10} (a), the big digit represented
in binary~(b), and the corresponding sequence
of small digits~(c).}
\label{fig:12}
\end{center}
\end{figure}

When $c$ is not a power of~2, we can no longer represent
a color drawn from $\{0,\ldots,c-1\}$ in $\log_2\! c$
bits, as in the standard representation of
Lemma~\ref{lem:atomic}.
Our core tool for coping with this complication is a
result of Dodis, P\v atra\c scu and Thorup~\cite{DodPT10}.
They demonstrate that a usual binary computer
can simulate a $C$-ary computer, for arbitrary
integer $C\ge 2$,
essentially without a loss in time or space in the
sense that an array of $n$ integers, each drawn from
$\{0,\ldots,C-1\}$, can be represented in
$n\log_2\! C+O(1)$ bits so as to support constant-time
reading and writing of individual array entries.
In order to be able to employ word parallelism, we use
this not with $C=c$, as would be most natural,
but with $C=c^m$ for some $m$ chosen essentially
to make $C\approx 2^w$.
Let us call elements of $\{0,\ldots,c-1\}$ and of
$\{0,\ldots,C-1\}$ \emph{small digits} and
\emph{big digits}, respectively.
A big digit is shown symbolically in
Fig.~\ref{fig:12}(a) as it is
represented in the data structure of
Dodis, P\v atra\c scu and Thorup;
it may be thought of a composed of $m$
$c$-ary digits, each of which is drawn as a triangle.
Once the big digit is read out of the data structure of
Dodis, P\v atra\c scu and Thorup, it is given by its
usual binary representation as a sequence of
$\Tceil{m\log_2\! c}$ bits.
In Fig.~\ref{fig:12}(b) each bit of the big digit
is shown as a dot.
If $c$ is not a power of~2, certain patterns of
values (namely those that represent the integers
$c^m,c^m+1,\ldots,2^{\Tceil{m\log c}}-1$) cannot occur.
This is symbolized in Fig.~\ref{fig:12}(b) by the
leftmost (most significant) dot being only
partially drawn;
put differently, the missing part of the
leftmost bit corresponds to a fraction of a bit 
that is wasted.
If we convert the big digit to the corresponding
sequence of small digits,
as shown in Fig.~\ref{fig:12}(c),
we can operate efficiently on the small digits
as on the standard representation in the proof
of Lemma~\ref{lem:atomic}.
Indeed, the sequence of small digits \emph{is}
in the standard representation of Lemma~\ref{lem:atomic},
only for a number of colors equal to
$2^f$, where $f=\Tceil{\log c}$---the largest colors
simply happen not to be present.
The conversion can be carried out with the
algorithm of Lemma~\ref{lem:base};
we shall express this by saying that we convert
the big digit
from base~$c$ to base $2^f$.

Just as when $c$ is a power of~2, we need a compact
representation that can be used to encode deficient
client vectors and essentially substitutes
base $c-1$ for base~$c$.
Since $c-1$ was never assumed to be a power of~2,
here the differences are small.
Once containers are available, the proof can proceed
as in Subsection~\ref{subsec:general2}.
We now describe the details and begin by providing
an analogue of Lemma~\ref{lem:atomic} for general
values of~$c$.

\begin{lemma}
\label{lem:atomicg}
There is an externally sized colored choice dictionary
$D$ that, for arbitrary given $m,c\in\TbbbN$,
can be initialized for universe size~$m$ and
$c$ colors in $O(\Tceil{{{(m\log c)}/w}})$ time
and subsequently needs to record in an
external bit whether its client vector is full,
stores its state as an
element of $\{0,\ldots,c^m-1\}$ and executes
\Tvn{color}, \Tvn{setcolor} and \Tvn{successor} in
$O(\Tceil{{{(m\log c)}/w}}\Lambda)$ time,
where $\Lambda=\log m+(\log(2+{{(m\log c)}/w}))^2$.
Moreover, at times when $D$'s client vector is
deficient, $D$'s state is bounded by
$2^{\Tfloor{m\log c-({m/{(2 c)}}+3 c)}}$.

Alternatively, if initialized with an additional
parameter $t\in\TbbbN$ and given access to certain tables
of at most $c^{m/t}$ bits that can be computed
in $O(c^{m/t})$ time and depend only on~$c$,
$m$ and~$t$,
$D$ can execute \Tvn{color}, \Tvn{setcolor}
and \Tvn{choice} in
$O(\Tceil{{{(m\log c)}/w}}t)$ time.

The transient space needed by $D$ is $O(w+m\log c)$ bits.
\end{lemma}

\begin{proof}
We again view $D$'s task as that
of maintaining a sequence
of $m$ color values or digits, each drawn from $\{0,\ldots,c-1\}$.
When $D$ is not in the compact representation,
it simply stores its state as the single $c$-ary
integer in $\{0,\ldots,c^m-1\}$ formed by the $m$ digits.
In order to execute
\Tvn{color}, \Tvn{setcolor} or \Tvn{successor},
$D$ converts its state from base~$c$ to base
$2^f$, where $f=\Tceil{\log c}$, obtaining what we call
the \emph{loose} representation, executes the operation
in question on the loose representation
as described in the proof of
Lemma~\ref{lem:atomic}, and
reconverts its state from base $2^f$ to base~$c$.
The base conversion is carried out with
the algorithm of
Lemma~\ref{lem:base} in
$O(\Tceil{{{(m\log c)}/w}}\Lambda)$ time,
and the operations executed on the loose representation
are no more expensive.

As in the proof of Lemma~\ref{lem:atomic},
in the compact representation a vector $Z$ of
$c$ bits is used to keep track of the set of
colors represented in~$D$.
When required to convert between the standard and
the compact representation, $D$ carries out the
conversion via the loose representation.
The conversion between the loose representation
and the compact representation is illustrated
in Fig.~\ref{fig:13}.
To go from the loose to the compact representation
at a time when the client vector is deficient
(Fig.~\ref{fig:13}(a)),
we first apply the excision operation to ensure that the
unused color is $c-1$ (Fig~\ref{fig:13}(b)).
Ignoring rounding issues, we proceed as follows:
Within groups, now of $2 c$ consecutive
digits each, we convert from base $2^f$ to base $c-1$
(Fig.~\ref{fig:13}(c)).
Since $2 c\log(c-1)\le 2 c\log c+2 c\ln(1-{1/c})\le
\Tfloor{2 c\log c}-1$, each group can be
stored in a field of $\Tfloor{2 c\log c}-1$
\emph{used} bits.
Appealing to the algorithm of
Lemma~\ref{lem:compaction}, we move the used bits of
all groups to
$({m/{(2 c)}})(\Tfloor{2 c\log c}-1)
\le\Tfloor{m\log c}-{m/{(2 c)}}$ consecutive
positions.
Since $\Tfloor{m\log c}$ unrestricted bits are available 
for storing the state of~$D$
(Fig.~\ref{fig:12}(b) has at least this many full dots),
this yields $m/{(2 c)}$ free bits.
Let us now take rounding into account.
The formation of groups may leave up to
$2 c-1$ digits that are not part of any group.
Even in the compact representation, we store
such digits to base $2^f$, which wastes less than
one bit per digit and at most $2 c$ digits altogether.
In addition, $c$ bits are occupied by~$Z$.
In summary, the number of free bits is at least
${m/{(2 c)}}-3 c$, as indicated in the lemma. 

\begin{figure}
\begin{center}
\epsffile{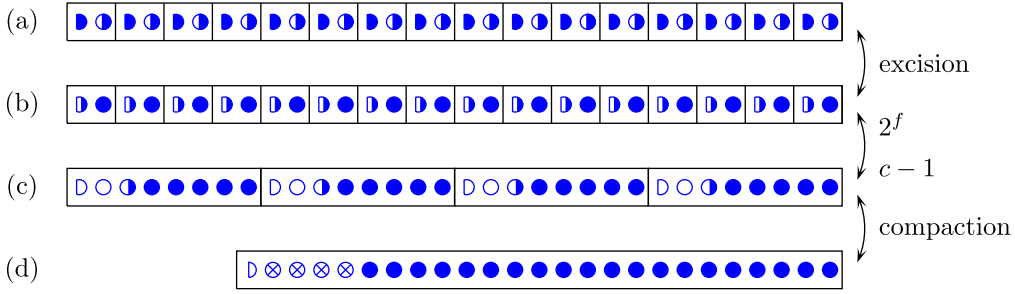}
\caption{The conversion between the loose and
the compact representation of a container.
A filled-in dot represents a bit that is in use,
a circle represents a free bit, and a circle
with a cross represents a bit that is
part of the payload.}
\label{fig:13}
\end{center}
\end{figure}

For given $t\in\TbbbN$,
suitable tables of at most $c^{m/t}$ bits that
can be computed in $O(c^{m/t})$ time and depend
only on $c$ and $m$ allow us to carry out all
base conversions in $O(\Tceil{{(m\log c)}/w}t)$ time,
which leads to the alternative time bounds.
\end{proof}

\subsection{An Unrestricted $c$-Color Choice Dictionary for General $c$}

In order to combine several instances of the
data structure of the previous lemma, we need
the data structure of
Dodis, P\v atra\c scu and Thorup~\cite{DodPT10}
as extended by Hagerup and Kammer~\cite{HagK16}
to support constant-time initialization.
Lemma~\ref{lem:dodis} below specializes
\cite[Theorem~6.5]{HagK16} to the case
$c_1=\cdots=c_{n-1}\ge c_n$.
On the other hand, Lemma~\ref{lem:dodis}
incorporates
a factor of $O(q\log q)$
not present in the formulation of~\cite{HagK16},
where $q=O(1)$ was assumed.
The factor accounts for the time needed to multiply
and divide integers of $O(q)$ $w$-bit words each
according to Lemma~\ref{lem:multiple}.

\begin{lemma}[\cite{HagK16}]
\label{lem:dodis}
There is a data structure
that, for all given $n,C,C'\in\TbbbN$
with $2\le C'\le C$,
can be initialized in constant time and subsequently
occupies $(n-1)\log_2\! C+\log_2\! C'+O((\log n)^2+1)$ bits
and maintains a sequence drawn from
$\{0,\ldots,C-1\}^{n-1}\times\{0,\ldots,C'-1\}$
under reading and writing
of individual elements in the sequence in
$O(q\log q)$ time, where $q=2+{{(\log C)}/w}$.
The data structure does not initialize the sequence.
The parameter $C$ may be presented
to the data structure in the form of a pair $(x,y)$
of positive integers with $C=x^y$ and
$y=n^{O(1)}$, and the analogous statement holds for~$C'$.
\end{lemma}

\begin{theorem}
\label{thm:colored}
There is an externally sized
choice dictionary
that, for arbitrary given $n,c\in\TbbbN$
with $c=n^{O(1)}$,
can be initialized for universe size $n$
and $c$ colors in constant time
and subsequently occupies $n\log_2\! c+O((\log n)^2+1)$ bits and
executes \Tvn{color}, \Tvn{setcolor}
and \Tvn{choice} in
$O(q(\log(c+\log n)+(\log q)^2))$ time,
where $q=2+{{c\log c(c+\log n)}/w}$.
In particular, if $c=O(w)$,
$q=O(1+c\log c)$ and
the operation times are
$O((\log\log(n+4)+(\log c)^2)c\log c)$.

Alternatively, if initialized with an additional
parameter $t\in\TbbbN$ and given access to suitable tables
of at most $c^{{c(c+\log n)}/t}$ bits that can be computed
in $O(c^{{c(c+\log n)}/t})$ time and depend only on $c$,
$n$ and~$t$,
$D$ can execute \Tvn{color}, \Tvn{setcolor} and
\Tvn{choice} in $O(q t)$ time.
If $c=O(w)$,
the operation times are $O(1+t c\log c)$.

The transient space needed by $D$ is $O(w+c(c+\log n)\log c)$ bits.
\end{theorem}

\begin{proof}
We use the same overall organization as in the data structure
of Theorem~\ref{thm:colored2}, but implement containers
via Lemma~\ref{lem:atomicg} and store the state of each
container as a ``big digit'' in a single global instance $D$
of the data structure of Lemma~\ref{lem:dodis}.
Recall that each container should offer a payload of
$\Theta(c+\log n)$ bits when its client vector is deficient.
We choose $m\in\TbbbN$ just large enough to meet this goal,
which for the data structure of Lemma~\ref{lem:atomicg}
means that $m=\Theta(c(c+\log n))$.
We then form $N=\Tfloor{n/m}$ containers, each with
a universe size of~$m$, and, if
$m$ does not divide $n$, an additional instance of the
data structure of Lemma~\ref{lem:atomicg} with a universe
size of $m\bmod n$ whose state is also stored in~$D$
but that is otherwise handled separately, similarly
to what happens to the surplus digits in the data
structure of Theorem~\ref{thm:colored2}.
The quantities $C=c^m$ and $C'=c^{m\bmod n}$
are available in the form $x^y$ with $x=c$ and
$y=n^{O(1)}$, as allowed by Lemma~\ref{lem:dodis}.

Each of the operations \Tvn{color}, \Tvn{setcolor}
and \Tvn{choice} carries out a constant number of
operations on containers.
To operate on a container, we first fetch its state
in $D$, which takes $O(q\log q)$ time according to
Lemma~\ref{lem:dodis}.
By Lemma~\ref{lem:atomicg}, the operation on the
container itself can be carried out in
$O(q\Lambda)$ time, where
$\Lambda=\log m+(\log q)^2=O(\log(c+\log n)+(\log q)^2)$.
Storing the new state of the container after the
operation in $D$ again takes $O(q\log q)$ time,
and an overall time bound of
$O(q(\log(c+\log n)+(\log q)^2))$ follows.
The alternative time bounds are obtained simply
by substituting $t$ for $\Lambda$.
\end{proof}

\section{Iteration}
\label{sec:iteration}

This section discusses iteration for the
choice dictionaries developed in the previous sections.
If a choice dictionary is to support iteration,
it must be supplied with additional space
(informally, for storing how far the iteration
has progressed), namely $O(\log n)$ bits
for an uncolored choice dictionary and
$O(c\log n)$ bits for a $c$-colored
choice dictionary, where $n$ is the universe size.
In the following discussion,
such additional space is assumed to be available.

\subsection{Iteration in the Uncolored Choice Dictionary}
\label{subsec:iter2}

When the data structure of Lemma~\ref{lem:main} is
used to realize an (uncolored) choice dictionary
with universe size $n$
and client set $S$, it is natural to associate
the $k$th component of the sequence
$(a_1,\ldots,a_N)$, for $k=1,\ldots,N$, with the
subset $U_k=\{2 b(k-1)+1,\ldots,2 b k\}$ of
the universe $U=\{1,\ldots,n\}$ and to view $a_k$
as a data structure that represents the
subset $S_k=S\cap U_k$ of $S$, each \texttt{1}
in the binary representation of $a_k$
corresponding to an element of~$S_k$.
With an algorithm of Lemma~\ref{lem:log}(a), it is easy to support
robust iteration over $S_k$ in constant time:
If $S_k$ is nonempty when $\Tvn{iterate}.\Tvn{next}$
is first called, enumerate $\min S_k$.
At every subsequent call of $\Tvn{iterate}.\Tvn{next}$,
enumerate the smallest element of $S_k$ larger than
the element most recently enumerated, ending the
enumeration when there is no such larger element.
Similarly, the dictionary $D_2$ of
Subsection~\ref{subsec:reduction} supports
constant-time robust iteration and can be
ignored in what follows.

By the considerations of the previous paragraph,
iteration in the choice dictionary of Section~\ref{sec:uncolored}
essentially boils down to iterating over those
$k\in V=\{1,\ldots,N\}$ with $a_k\not=0$, i.e., over
the strong $k\in V$.
Noting that the set of strong $k\in V$ at all times
is $\{\Tvn{mate}(\ell)\mid\ell\in\{\mu+1,\ldots,N\}\}$,
it is easy to achieve this in the 
static case, i.e., when no insertions or
deletions take place during the iteration:
For $\eta=N,N-1,\ldots,\mu+1$, enumerate $\Tvn{mate}(\eta)$.
We think of $\eta$ as carrying out a sweep
from $N$ down to $\mu+1$, always enumerating
$\Tvn{mate}(\ell)$ for each $\ell$ encountered.
Even in the general (not necessarily static)
case, ideally, we should have
$R_0\subseteq L\subseteq R$, where $R$ is the set
of strong integers in $V$ that
have not been enumerated,
$R_0\subseteq R$ is the set of such integers that
have been strong since the beginning of the iteration
and $L=\{\Tvn{mate}(\ell)\mid\ell\in\{\mu+1,\ldots,\eta\}\}$.
A strong integer $k\in V$ is considered to belong to $R$
and possibly $R_0$ as long as not all elements
represented by $a_k$ have been enumerated.

A deletion may cause an element $k$ of $R$
to the left of the barrier
to drop out of $L$, namely if $k$ swaps a 
mate in front of the sweep (and therefore still to be swept over)
for a mate behind the sweep
(this may happen in the situations of
Fig.\ \ref{fig:4}(c) and Fig.~\ref{fig:5}(c)).
In order to ``rescue'' such elements $k$,
we collect them in a set $S^+$ and
iterate also over $S^+$, as described below.
On the other hand, an insertion may cause
an element to the left of the barrier
that has already been enumerated and left
$R$ and $L$ at that time to reenter $L$ by acquiring a
new mate in front of the sweep
(see Figs.\ \ref{fig:3}, parts (a) and~(c)).
In an attempt to prevent such elements from
being enumerated again, we store them in
a set $S^-$ of elements to be skipped.
The details follow.

We introduce two dynamic
uncolored choice dictionaries distributed
over $A[1],\ldots,A[N]$, $D^+$ with client set $S^+$
and $D^-$ with client set $S^-$.
As in
the proof of Theorem~\ref{thm:colored2},
$D^+$
and $D^-$ are realized via bits in the auxiliary
fields, and their universe sizes are kept equal
to $\mu$ at all times.
By means of a buffer of $\Tceil{\log(n+1)}$ bits,
initialized to the value~0,
we ensure that if $D^+.\Tvn{choice}$ at some
point returns a nonzero integer $k$, subsequent
calls of $D^+.\Tvn{choice}$ will return
the same integer~$k$
for as long as $k$ remains an element of~$S^+$.
The method is simple:
When the value of the buffer is~0, the integer
returned by a call of $D^+.\Tvn{choice}$ is
also stored in the buffer, and as long as the value
of the buffer remains nonzero, subsequent calls
of $D^+.\Tvn{choice}$ simply return that value
instead of executing the normal steps.
Finally, when the value in the buffer is
deleted from $D^+$, the buffer is set to~0.
This mechanism helps to ensure that once the
enumeration of the elements represented by an
integer $a_k$ has started, where $k\in V$,
it is completed without
intervening enumeration of other elements.
The iteration also uses an integer $\eta\in\{0,\ldots,n\}$
to keep track of the sweep.

To start an iteration (i.e., to execute
$\Tvn{iterate}.\Tvn{init}$), set $\eta:=N$
and initialize $D^+$ and $D^-$ (to
$S^+=S^-=\emptyset$) for a
universe size of (the current value of) $\mu$.
To test whether elements remain to be
enumerated (i.e., to execute
$\Tvn{iterate}.\Tvn{more}$), test whether
$S^+\not=\emptyset$ or $\eta>\mu$.
Finally, to enumerate the next element
(i.e., to execute $\Tvn{iterate}.\Tvn{next}$),
do the following:
If $S^+\not=\emptyset$, let $k=D^+.\Tvn{choice}$
and enumerate the next element of $a_k$.
If subsequently $a_k$ has no more elements
to be enumerated, delete $k$ from $S^+$.
If instead $S^+=\emptyset$ but $\eta>\mu$,
let $k=\Tvn{mate}(\eta)$.
If $k\not\in S^-$, proceed to enumerate
the next element of $a_k$.
If subsequently $a_k$ has no more elements
to be enumerated, decrease $\eta$ by~1.
If $k\in S^-$, also decrease $\eta$ by~1.
Since no element was enumerated in this case, however,
restart $\Tvn{iterate}.\Tvn{next}$ recursively.

When an insertion or a deletion causes an
element of $V\setminus S^-$ to drop out of $L$ even
though it continues to be strong,
insert it in $D^+$---it must be to the
left of the barrier.
When a deletion causes an element of $S^+$ to
become weak, delete it from $D^+$.
Finally, when an insertion causes an element of $V$
to the left of the barrier
to enter $L$ even though it was not in
$L\cup S^+$ before the insertion,
insert it in $D^-$;
an inspection of Figs.\ \ref{fig:4} and~\ref{fig:5}
shows that this cannot happen in a deletion.
In addition to these situations in which $S^+$
and $S^-$ are explicitly changed, one may note
that if an insertion causes an integer $k$
to leave $\{1,\ldots,\mu\}$ ($k$ crosses the barrier
from left to right), then $k$ automatically
drops out of $S^+$ and $S^-$.

It can be verified by induction that between operations,
the following holds at all times during an iteration:
$L\setminus S^-$ and $S^+$ are disjoint and contain
only strong integers.
Moreover,
$R_0\subseteq (L\setminus S^-)\cup S^+$,
so that all elements that should be enumerated
are actually enumerated.
In the \emph{decremental} case, i.e., when there are no
insertions during an iteration, $S^-$ is always
empty, and the stronger invariant
$R=R_0=L\cup S^+$ holds,
As a consequence, the iteration can be seen
to be robust.

In the \emph{incremental} case, i.e., when there
are no deletions during an iteration, an element
$k\in V$ may drop out of $S^-$ implicitly,
namely by virtue of crossing the barrier,
as explained above.
Informally, this causes the data structure to lose
knowledge of the fact that $k$ was already enumerated,
and as a result $k$ may be enumerated a second time.
However, because the barrier moves only in one direction,
this can happen at most once to each $k\in V$, so
that no element is enumerated more than twice.
The iteration no longer happens in constant
worst-case time because of the need to skip
elements of $L\cap S^-$.
Skipping an element $k\in L\cap S^-$ takes
constant time and removes $k$ from $L$.
Even though $\Tvn{iterate}.\Tvn{more}$ is only a
query operation, it should update the state of the
iteration to prevent elements from being
skipped repeatedly.
Since $L\cap S^-$ is empty at the beginning of
an iteration and no
insertion causes more than a constant number
of integers to enter $L\cap S^-$, the total time
wasted in skipping elements of $L\cap S^-$
up to a certain point of an iteration can then be seen
to be within a constant factor of the number
of insertions carried out since the beginning
of the iteration.

In the general case, when insertions and deletions may
be arbitrarily intermingled, we can still bound
the time spent in skipping elements of $L\cap S^-$
by a constant times $O(n\Tsub u)$, where $n\Tsub u$
is the number of insertions and
deletions carried out since the beginning of
the iteration.
The number of times that an element $k\in V$
is enumerated a second or later time is also
$O(n\Tsub u)$, and each such enumeration causes
at most $2 b$ elements represented by $a_k$ to
be enumerated again.
Since we can choose $b=\Theta(\log n)$, this
leads to a bound of $O(n\Tsub u\log n)$ on the
time spent in unwanted enumerations.

The results of this subsection can be summarized
as follows:

\begin{theorem}
There is a self-contained (uncolored) choice dictionary
that, for arbitrary $n\in\TbbbN$, can be initialized for
universe size $n$ in constant time and subsequently
occupies $n+O(\log n)$ bits, executes $\Tvn{insert}$,
\Tvn{delete}, \Tvn{contains} and \Tvn{choice} in
constant time and supports the following
forms of iteration:
If during an iteration there are no calls of
\Tvn{insert} (the decremental case), the iteration
is robust and works in constant time.
If during an iteration there are no calls of
\Tvn{delete} (the incremental case), the
iteration is robust, except that each integer
that is enumerated may be enumerated a second time,
and the iteration works in constant amortized time in the
following sense:
At a time when, since the start of an ongoing iteration,
there has been $n\Tsub s$ calls of $\Tvn{iterate}.\Tvn{next}$
and $\Tvn{iterate}.\Tvn{more}$
and $n\Tsub i$ calls of \Tvn{insert}, the total time
spent in the single call of $\Tvn{iterate}.\Tvn{init}$
and in the $n\Tsub s$ calls of $\Tvn{iterate}.\Tvn{next}$
and $\Tvn{iterate}.\Tvn{more}$ is
$O(1+n\Tsub s+n\Tsub i)$.
If during an iteration insertions and deletions may
be carried out in an arbitrary order, finally, only elements
of the client set are enumerated, but an integer may
be enumerated repeatedly.
If an iteration starts with $n_0$ elements in the
client set and
there are
$n\Tsub u$ calls of \Tvn{insert} and \Tvn{delete} during
a period of time from the start of the iteration,
the total time
spent in the single call of $\Tvn{iterate}.\Tvn{init}$
and in calls of $\Tvn{iterate}.\Tvn{next}$
and $\Tvn{iterate}.\Tvn{more}$ during that period of time
is $O(1+n_0+n\Tsub u\log n)$.
\end{theorem}

\subsection{Iteration in the Colored Choice Dictionaries}

For the colored choice dictionaries
of Sections~\ref{sec:colored2} and \ref{sec:colored},
we can support efficient iteration only if each
iteration is \emph{complete}, i.e.,
is allowed
to enumerate all elements without being terminated early.
Since complete iterations are common, this
still leaves interesting applications.
Complete iteration can be added to the choice dictionaries
of Theorems \ref{thm:colored2},
\ref{thm:colored2t} and \ref{thm:colored}.
We give only one example, corresponding to
Theorem~\ref{thm:colored}, and specialize
the theorem to the case of constant $c$.

\begin{theorem}
\label{thm:iteration}
There is a self-contained choice dictionary $D$ that,
for arbitrary given $n\in\TbbbN$ and constant
$c\in\TbbbN$, can be initialized for universe
size~$n$ and $c$ colors in constant time
and subsequently occupies
$n\log_2\! c+O((\log n)^2+1)$ bits,
executes \Tvn{color}, \Tvn{setcolor} and
\Tvn{choice} in $O(\log\log n)$ time and,
for $j=0,\ldots,c-1$, supports complete iteration over
$S_j$, where the client vector of $D$ is
$(S_0,\ldots,S_{c-1})$, as follows:
Only elements of $S_j$ are enumerated, every integer
that belongs to $S_j$ during the entire iteration
is enumerated, but an
integer may be enumerated repeatedly.
If an iteration over $S_j$ starts with $|S_j|=n_j$
and there 
are $n\Tsub u$ calls of \Tvn{setcolor} during a period
of time from the start of the iteration,
the total time
spent in the single call of $\Tvn{iterate}.\Tvn{init}$
and in calls of $\Tvn{iterate}.\Tvn{next}$
and $\Tvn{iterate}.\Tvn{more}$ during that period of time
is $O(1+n_j\log\log n+n\Tsub u\log n)$.

Alternatively, if initialized with an additional
parameter $t\in\TbbbN$ and given access to suitable
tables of at most $n^{1/t}$ bits that can be computed
in $O(n^{1/t})$ time and depend only on $n$,
$c$ and $t$, $D$ can execute \Tvn{color},
\Tvn{setcolor} and \Tvn{choice} in
$O(t)$ time, and the time bound above for iteration
is replaced by $O(1+n_j t+n\Tsub u\log n)$.
\end{theorem}

\begin{proof}
Let the client vector of $D$ be $(S_0,\ldots,S_{c-1})$.
With notation as in the proof of Theorem~\ref{thm:colored2},
we carry out the following procedure to iterate
over $S_j$:
For $\eta=N,N-1,\ldots,\mu$, enumerate both
$\eta$ and $\Tvn{mate}(\eta)$, where enumerating an
element $k$ of $V=\{1,\ldots,N\}$ now means enumerating all
occurrences of the color $j$ in $a_k$.
This enumerates all occurrences of $j$ to the right
of the barrier at least once.
In order to enumerate the occurrences of $j$ to the
left of the barrier, we use the dynamic uncolored
dictionary $D_j$ of the proof of Theorem~\ref{thm:colored2},
but split it into two, $D'_j$ and $D''_j$.
At the beginning of each iteration, one of
$D'_j$ and $D''_j$ is empty---here
we need the previous iteration to have been complete.
Suppose that $D'_j$ is nonempty at the beginning
of the iteration.
Then we repeatedly let $k:=D'_j.\Tvn{choice}$ and,
as long as $k\not=0$,
enumerate $k$, delete $k$ from $D'_j$ and insert
$k$ in $D''_j$.
In the next iteration, $D'_j$ and $D''_j$
switch roles.
The other uses of $D_j$ can easily be adjusted to the
fact that the data structure now consists of two
components.
If there are no calls of \Tvn{setcolor} during an
iteration, the iteration is robust, and the iteration
takes $O(\log\log n)$ or $O(t)$ time like the other
operations on~$D$.
Each call of \Tvn{setcolor} adds a constant number
of containers whose occurrences of the color $j$
may be enumerated again.
This needs $O(\log\log n)$ or $O(t)$ time once for
converting the container to the loose representation
plus constant time for each of $O(\log n)$ elements
enumerated, for a total time of $O(\log n)$
(without loss of generality $t=O(\log n)$).
This shows the time bounds claimed for iteration.
\end{proof}

\subsection{Breadth-First Search and Shortest-Path Forests}

An application of Theorem~\ref{thm:iteration}
is to breadth-first search (BFS) and the computation
of shortest-path forests.
The following more precise definitions
were adapted from~\cite{HagK16}.

Given a directed or undirected
$n$-vertex graph $G=(V,E)$
and a permutation 
$\pi$ of $V$, i.e., a
bijection from $\{1,\ldots,n\}$ to~$V$,
we define a \emph{spanning forest} of $G$
\emph{consistent with} $\pi$ to be a sequence
$F=(T_1,\ldots,T_q)$, where $T_1,\ldots,T_q$
are vertex-disjoint outtrees that are
subgraphs of $G$ (if $G$ is directed)
or of the directed version of~$G$
(if $G$ is undirected)
and the union of whose vertex sets
is $V$, such that for each $v\in V$, the root of
the tree in $\{T_1,\ldots,T_q\}$ that contains~$v$
is the first vertex in the sequence
$(\pi(1),\ldots,\pi(n))$ from which $v$ is reachable in~$G$.
If, in addition, every path in the union
of $T_1,\ldots,T_q$
(if $G$ is directed)
or its undirected version (if $G$ is undirected)
is a shortest path in $G$, $F$ is a
\emph{shortest-path spanning forest} of $G$
consistent with~$\pi$.
A shortest-path spanning forest of $G$ can be
produced by a BFS that,
whenever there are no vertices adjacent
to those already processed, picks the next
vertex to process as the first vertex, in
the order given by $\pi$, that has not yet
been processed.

By computing a shortest-path
spanning forest $F=(T_1,\ldots,T_q)$
of an $n$-vertex graph $G=(V,E)$ consistent with
a permutation $\pi$ of $G$ we mean producing a
sequence $((u_1,v_1,k_1,d_1),\ldots,$\break$(u_n,v_n,k_n,d_n))$ of
4-tuples with $u_i\in V\cup\{0\}$, $v_i\in V$,
$k_i\in\TbbbN$ and $d_i\in\TbbbN_0$
for $i=1,\ldots,n$ such that
$k_1\le\cdots\le k_n$,
$\{v_i\mid 1\le i\le n$ and $k_i=j\}$ and
$\{(u_i,v_i)\mid 1\le i\le n$,
$k_i=j$ and $u_i\not=0\}$ 
are precisely the vertex and edge sets of~$T_j$,
respectively, for $j=1,\ldots,q$,
and $d_i$ is the depth of $v_i$ in $T_{k_i}$,
for $i=1,\ldots,n$.
If, in addition, for each $\ell\in\{1,\ldots,n\}$
with $u_\ell\not=0$ there is an $i\in\{1,\ldots,\ell-1\}$
with $v_i=u_\ell$, we say that $F$ is computed in
\emph{top-down order}.
Thus for $j=1,\ldots,q$, the root and
the edges of $T_j$ are to
be output (in a top-down order),
each with the index $j$ of its tree~$T_j$
and an indication of the depth in~$T_j$.

Hagerup and Kammer describe an algorithm for
computing a shortest-path spanning forest of
a given graph $G=(V,E)$ that stores a color
drawn from $\{\mbox{white},\mbox{gray},\mbox{black}\}$
for each vertex in $V$ in a 3-color choice dictionary
$D$ and needs linear time outside of a number
of complete iterations over the set of
gray vertices \cite[Theorem 8.5]{HagK16}.
Suppose that $G$ has $n$ vertices and $m$ edges.
As is easy to see from the proof, the sum, over
all iterations, of the number of gray vertices
present at the beginning of the iteration
is $O(n)$, and repeated enumerations of a gray vertex
do not jeopardize the correctness of the algorithm.
Therefore Theorem~\ref{thm:iteration} implies
the following new result.

\begin{theorem}
\label{thm:bfs}
Given a directed or undirected graph $G=(V,E)$
with $n$ vertices and $m$ edges and
a permutation $\pi$ of $V$,
a shortest-path
spanning forest of $G$ consistent with $\pi$
can be computed in
top-down order in $O(n\log n+m\log\log n)$ time with
$n\log_2\!3+O((\log n)^2+1)$ bits of working memory.
Alternatively, for every given $t\in\TbbbN$,
the problem can be solved in
$O(n\log n+m t)$ time with
$n\log_2\! 3+O(n^{1/t}+(\log n)^2)$ bits.
If $G$ is directed, its representation must
allow iteration over the inneighbors and
outneighbors of a given
vertex in time proportional to their number
plus a constant
(in the terminology
of~\cite{ElmHK15},
$G$ must be given with in/out adjacency lists).
\end{theorem}

\bibliography{all}

\end{document}